\catcode`\@=11

\newif\if@fewtab\@fewtabtrue
{\count255=\time\divide\count255 by 60
\xdef\hourmin{\number\count255}
\multiply\count255 by-60\advance\count255 by\time
\xdef\hourmin{\hourmin:\ifnum\count255<10 0\fi\the\count255}
\def\ps@draft{\let\@mkboth\@gobbletwo
    \def\@oddhead{}
    \def\@oddfoot{\hbox to 7 cm{\tiny \versionno
       \hfil}\hskip -7cm\hfil\rm\thepage \hfil}
    \def\@evenhead{}\let\@evenfoot\@oddfoot}

\def\draftcite#1{\ifnum\draftcontrol=1#1\else{}\fi}
\def\@lbibitem[#1]#2{\item{}\hskip -3cm \hbox to 2cm
{\hfil$\scriptstyle\draftcite{#2}$}\hskip
1cm[\@biblabel{#1}]\if@filesw
     {\def\protect##1{\string ##1\space}\immediate
      \write\@auxout{\string\bibcite{#2}{#1}}}\fi\ignorespaces}

\def\@bibitem#1{\item\hskip -3cm \hbox to 2cm
{\hfil {\footnotesize\draftcite{#1}}}\hskip 1cm
\if@filesw \immediate\write\@auxout
       {\string\bibcite{#1}{\the\value{\@listctr}}}\fi\ignorespaces}

\def\citen#1{\if@filesw \immediate\write \@auxout {\string\citation{#1}}\fi%
\@tempcntb\m@ne \let\@h@ld\relax \def\@citea{}%
\@for \@citeb:=#1\do {\@ifundefined {b@\@citeb}%
    {\@h@ld\@citea\@tempcntb\m@ne{\bf ?}%
    \@warning {Citation `\@citeb ' on page \thepage \space undefined}}%
    {\@tempcnta\@tempcntb \advance\@tempcnta\@ne
    \setbox\z@\hbox\bgroup\ifcat0\csname b@\@citeb \endcsname \relax
    \egroup \@tempcntb\number\csname b@\@citeb \endcsname \relax
    \else \egroup \@tempcntb\m@ne \fi \ifnum\@tempcnta=\@tempcntb
    \ifx\@h@ld\relax \edef \@h@ld{\@citea\csname b@\@citeb\endcsname}%
    \else \edef\@h@ld{\hbox{--}\penalty\@highpenalty
    \csname b@\@citeb\endcsname}\fi
    \else \@h@ld\@citea\csname b@\@citeb \endcsname \let\@h@ld\relax \fi}%
\def\@citea{,\penalty\@highpenalty\hskip.13em plus.13em minus.13em}}\@h@ld}
\def\@citex[#1]#2{\@cite{\citen{#2}}{#1}}%
\def\@cite#1#2{\leavevmode\unskip\ifnum\lastpenalty=\z@\penalty\@highpenalty\fi%
  \ [{\multiply\@highpenalty 3 #1%
  \if@tempswa,\penalty\@highpenalty\ #2\fi}]}   %
\makeatother 

\def\aff           {affine Lie algebra}
\newcommand\Ahr[2] {\hat A^{(#1)}(#2)}
\newcommand\Adhr[2] {\hat A_{\Delta}^{(#1)}(#2)}
\def\bfe           {{\bf1}}
\def\be            {\begin{equation}}
\newcommand\bhr[2]	{\hat b^{(#1)}(#2)}
\newcommand\bra[1] 	{\langle \, #1 \mid}

\def\cala          {{\cal A}}
\def\calao         {{\cal A}^{(0)}}
\def\calap         {{\cal A}^{(+)}}
\def\calam         {{\cal A}^{(-)}}

\def\calc          {{\cal C}}

\def\call          {{\cal L}}

\def\caln          {{\cal N}}

\def\calS          {{\cal S}}

\def\cft           {conformal field theory}
\def\cfts          {conformal field theories}
\def\ch            {\hat c}
\def\Chi           {{\cal X}}

\def\chidihh            {\hat{\chi}_{\Delta_i}}
\newcommand\chidihhr[1] {\hat{\chi}^{(#1)}_{\Delta_i}}

\newcommand\chidn[1]	{\chi_{\Delta_#1}}

\def\Deltah        {\hat{\Delta}}
\def\delz          {\partial_z}

\def\dl            {\mathbb }

\def\drac          {\displaystyle\frac }
\def\dsum          {\displaystyle\sum}

\newcommand\Eahr[3]	{\hat E_{#1}^{(#2)}(#3)}
\def\ee            {\end{equation}}
\def\eE            {{\rm e}}

\newcommand\erf[1]{(\ref{#1})}
\def\findim        {finite-dimensional}
\newcommand{\fline}[1]{\vfill\noindent ------------------\\[1 mm]}

\def\futnot#1      {\ifnum\draftcontrol=1
                   \footnote{~{\sc internal footnote:} #1}\ \fi}
\def\futnote#1     {\footnote{~#1}\ }
\def\g             {{\sf g}}

\newcommand\Gr[2]  {\hat G^{(#1)} (#2)}
\def\h             {{\sf h}}

\def\hil           {{\cal H}}
\newcommand\Hhhr[1] 	{\hat H^{(#1)}}

\def\hy            {$\mbox{-\hspace{-.66 mm}-}$}

\def\ii            {{\rm i}}

\def\infdim        {infinite-dimensional}

\def\irrep         {irreducible representation}
\def\J             {J}
\newcommand\Jahr[3]	{\hat J_{#1}^{(#2)}(#3)}
\newcommand\Jahhr[2]	{\hat {J}_{#1}^{(#2)}}

\newcommand\Jr[2]  {\hat J^{(#1)}(#2)}
\def\kma           {Kac\hy Moo\-dy algebra}

\long\def\labl#1   {\label{#1}\ee \ifnum\draftcontrol=1
                   \mbox{ }\\[-12 mm]\query{#1}\\[5 mm] \fi}
\def\lambdam       {\lambda}

\newcommand\lamhr[2]	{\hat\Lambda^{(#1)}(#2)}

\def\lie           {Lie algebra}

\newcommand\Lv[1]  {L(#1)}

\newcommand\Lvr[2]  {\hat L^{(#1)} (#2)}

\newcommand\mket[1]{ \mid #1 \, \rangle} 

\def\modinv        {modular invarian}
\def\mono          {\eE^{2\pi\ii}}

\def\nontriv       {non-trivial}
\def\nord		{:}

\def\phid          {\varphi_\Delta}

\newcommand\phidhhr[1] {\hat{\varphi}^{(#1)}_\Delta}

\def\phidhh        {\hat{\varphi}_\Delta}
\newcommand\phidn[1]	{\varphi_{\Delta_#1}}
\def\phidohh		{\hat{\varphi}_{\Delta_1}}   
\newcommand\phidohhr[1] {\hat{\varphi}^{(#1)}_{\Delta_1}}
\newcommand\phidthhr[1] {\hat{\varphi}^{(#1)}_{\Delta_2}}

\newcommand\phidm[1]  {\varphi_\Delta (#1)}
\newcommand\phidhmr[2] {\hat\varphi^{(#1)}_\Delta(#2)}
\newcommand\phidnhhr[2] {\hat{\varphi}^{(#2)}_{\Delta_#1}}
\newcommand\phidnmohhr[1] {\hat{\varphi}^{(#1)}_{\Delta_{n-1}}}
\def\phidthh 		{\hat{\varphi}_{\Delta_2}}
\newcommand\phivhhr[1] {\hat{\varphi}^{(#1)}_0}
\newcommand\phivhmr[2] {\hat\varphi^{(#1)}_0(#2)}

\long\def\query#1{\hskip 0pt{\vadjust{\everypar={}\small\vtop to 0pt{\hbox{}%
     \vskip -13pt\rlap{\hbox to 50.0pc{\hfil{\vtop{\hsize=8pc\tolerance=6000%
     \hfuzz=.5pc\rightskip=0pt plus 3em\noindent#1}}}}\vss}}}}%

\def\reals         {\mbox{${\dl R}$}}
\def\reg           {{\rm reg}}
\def\rep           {representation}
\def\Rep           {Representation}

\newcommand\sect[1] {\section{#1}\setcounter{equation}{0}}
\newcommand\Sect[2]{\sect{#1}\label{s.#2} \ifnum\draftcontrol=1 \query{s.#2}\fi}

\def\SltR          {\mbox{SL(2,\reals)}}

\def\Thh           {\hat{ T}}
\newcommand\Thhr[1] {\hat{T}^{(#1)}}
\def\Tr            {{\rm Tr}}
\def\twodim        {two-di\-men\-si\-o\-nal}

\newcommand\version[1] {\ifnum\draftcontrol=1 \typeout{}\typeout{#1}\typeout{}
                   \vskip3mm \centerline{\fbox{\tt DRAFT -- #1 -- \today}}  
                   \vskip3mm \fi}

\newcommand\Whr[2] 	{\hat W^{(#1)}(#2)}
\newcommand\Widetilde[1] {\widehat{#1}}

\def\Wt                 {\mbox{${\rm W}_3$}}

\def\wzwm          {WZW model}

\def\zet           {{\dl Z}}

\global\def\draftcontrol{0}
\catcode`\@=12

\documentclass[12pt]{article} \usepackage{amssymb,amsfonts}

\setlength{\textwidth}{17cm} \setlength{\textheight}{22cm}
\hoffset -22mm \topmargin= -13mm \raggedbottom

\begin{document} 

\begin{flushright}  {~} \\[-23 mm] {\sf hep-th/9701061} \\
{\sf UCB-PTH-97/02} \\ {\sf LBNL-39809} 
\\[1 mm]{\sf January 1997} \end{flushright} \vskip 2mm

\begin{center} \vskip 12mm

{\Large\bf SYSTEMATIC APPROACH TO} \vskip 0.3cm
{\Large\bf CYCLIC ORBIFOLDS} \vskip 12mm
{\large L.\ Borisov} $^1$\, , \, {\large M.B.\ Halpern}$^2$\, , \,
{\large C.\ Schweigert}$^2$ \\[9mm] 
{$^1$ \small Mathematical Sciences Research Institute}\\
{\small 1000 Centennial 
Drive, Berkeley, CA 94720, USA} \\[2mm] {$^2$ \small Department of Physics,
University of California}\\
{\small and} \\
{\small Theoretical Physics Group, Lawrence Berkeley National Laboratory}\\ 
{\small Berkeley, CA 94720, USA}
\end{center}

\vskip 15mm

\begin{quote} {\bf Abstract} \\
We introduce an orbifold induction procedure  which provides a systematic
construction of cyclic orbifolds, including their twisted sectors.
The procedure gives counterparts in the orbifold theory of all the 
current-algebraic constructions of conformal field theory and enables us to 
find the orbifold characters and their modular transformation properties. 
\end{quote}

\vskip 12mm

\sect{Introduction}

Twisted scalar fields \cite{hath,sieg} and twisted vertex operators 
\cite{cofa,lewi} were introduced in 1971 and 1975 respectively, 
although their role in string theory was not fully understood
until the theory of orbifolds 
\cite{frlm,dhvw,dhvw2,dmfs,hava,dvvv,klsC,fukS} 
was developed in the mid 1980's.

In this paper, we focus on the class of {\em cyclic orbifolds}, which are
those  formed by modding out the $\zet_\lambdam$ symmetry (cyclic permutations)
of $\lambdam$ copies of a mother \cft. In particular, we give an
{\em orbifold induction procedure} which generates the twisted sectors of
these orbifolds directly from the mother theory,
\be\begin{array}{rcl}
\mbox{(mother) CFT}&\to&\mbox{(cyclic orbifold)}_\lambdam \, \\[.5pt]
&\lambdam&\end{array}\ee
without having to consider the tensor product theory. Thus, the orbifold
induction procedure makes what is apparently the hardest part of the problem 
into the easiest part.

Starting on the sphere (Sections 2 and 3), we find first that the orbifold 
induction procedure
generates a new class of \infdim\ algebras which we call the 
{\em orbifold algebras}. These algebras, which are operative in the
twisted sectors of cyclic orbifolds, are the root-covering algebras of the 
\infdim\ algebras of
\cft. Some subalgebras of these orbifold algebras appear earlier
in Refs.\ \cite{kawa5,fukS,bouw6}, and the role of these subalgebras in
cyclic orbifolds was emphasized 
by Fuchs, Klemm and Schmidt \cite{fukS}. With the help of the orbifold 
algebras, we find that all the familiar constructions of \cft\ have 
their counterparts in the orbifolds, including 
affine-Sugawara constructions \cite{baha,halp,dafr,knza}, 
coset constructions \cite{baha,halp,goko,goko2} and 
affine-Virasoro constructions \cite{haki,mprst,Hkoc}, as well as
$\SltR$-Ward identities \cite{bepz} and null-state
differential equations of BPZ type.
Higher-level counterparts of the
vertex operator constructions \cite{halp3,halp6,bahn,lewi,frka,sega}
and conformal embeddings \cite{baha,halp3,gono,scwa,babo} are also found.

The partition functions of cyclic orbifolds have been given for prime 
$\lambdam$ by Klemm and Schmidt \cite{klsC}. The orbifold induction procedure 
on the torus (Section 4) allows us to go beyond the partition
functions, giving 
a systematic description of
the characters of the orbifold, including their
modular properties and fusion rules. 

\sect{Orbifold Algebras}

The {\it orbifold algebras} , simple examples of which are displayed in 
mode form below, are the root-covering algebras of the infinite-dimensional 
algebras of conformal field theory. The orbifold algebras operate in the 
twisted sectors of cyclic orbifolds and these algebras may be induced in the 
orbifold by the {\it orbifold induction procedure} (see Subsection 2.6) 
from the infinite-dimensional algebras of the mother \cft. The identity 
of the orbifold algebras as root-covering algebras will be clear in 
Section 3, where we express the induction in its local form. 
The origin of the orbifold algebras is discussed from a more 
conventional viewpoint of cyclic orbifolds in Subsection 3.8.
In what follows,
the positive integer $\lambdam$ is the order of the cyclic orbifold.

\subsection{Orbifold affine algebra}

The simplest examples of orbifold algebras are the {\it orbifold affine 
algebras}, whose generators are 
\be \Jahr a r{m+\frac r\lambdam},\,\,\,\,\,a=1\ldots\dim\g,\,\,\,
\,\,r=0\ldots \lambdam-1,\,\,\,\,\,m\in\zet \ee
where $\g$ is any simple \findim\ Lie algebra. These algebras have 
the form 
\be\begin{array}{lll} [\Jahr a r {m+\frac r\lambdam},\Jahr b s {n+\frac
s\lambdam}]&=& \ii f_{ab}^{\,\,\,\,c}\Jahr c {r+s} {m+n+\frac {r+s}\lambdam}
\\[1em] &&
+\hat k\eta_{ab}(m+\frac r\lambdam)\delta_{m+n+\frac {r+s}\lambdam,\,0}
\,  \end{array}\labl{2.1}
where $f_{ab}^{\,\,\,\,c}$ and $\eta_{ab}$ are respectively the structure 
constants and Killing metric of $\g$, and $\hat k$ is the level of the 
orbifold affine algebra. 
The algebra \erf{2.1} is understood with the periodicity condition
\be \Jahr a {r\pm\lambdam} {m+\frac {r\pm\lambdam}\lambdam}=
\Jahr a r {m\pm1+\frac r\lambdam}.\ee
This is a special case (with $\epsilon=0$) of the general periodicity 
condition 
\be \Ahr {r\pm\lambdam} {m+\frac {r\pm\lambdam+\epsilon}\lambdam}=
\Ahr r {m\pm1+\frac {r+\epsilon}\lambdam}\labl{periodicity}
which will be assumed to hold with $\epsilon=0$ or $\frac12$ for the 
modes of all the orbifold algebras below.
It will also be useful to have the relation
\be \Ahr {-r} {m+\frac {-r+\epsilon}\lambdam}=\Ahr {\lambdam-r}
{m-1+\frac{\lambdam-r+\epsilon}\lambdam}\labl{-r}
which is a consequence of the periodicity \erf{periodicity}.

\subsection{Orbifold Virasoro algebra}

The generators of the {\it orbifold Virasoro algebras} are
\be \Lvr r {m+\frac r \lambdam},\,\,\,r=0\ldots\lambdam -1,
\,\,\,m\in\zet\labl{new}
which satisfy 
\be\begin{array}{ll}
[\Lvr r {m+\frac r \lambdam},\Lvr s {n+\frac s \lambdam}]=&
(m-n+\frac{r-s}\lambdam)\Lvr {r+s} {m+n+\frac{r+s}\lambdam}\\[1em]&+
\frac{\hat c}{12}\, (m+\frac r\lambdam)((m+\frac r\lambdam)^2-1)
\delta_{m+n+\frac{r+s}\lambdam,\,0}\end{array}\labl{2.5}
and the periodicity condition \erf{periodicity} with $\epsilon=0$. 
The quantity $\hat c\,$ is the central charge of the orbifold
Virasoro algebra. 

\subsection{Orbifold N=1 superconformal algebra}
\underline{\bf  $\widehat{NS}$ sector}\\[2em]
In the $\widehat{NS}$ sector of the N=1 orbifold superconformal algebra, we 
have, in addition to the orbifold Virasoro generators \erf{new}, the set
of $\lambdam$ ${}$ $\widehat{NS}$ orbifold supercurrents
\be \Gr r {m+\frac {r+\frac12}\lambdam},\,\,\,r=0\ldots\lambdam-1,
\,\,\,m\in\zet\ee
which satisfy the algebra
\be\begin{array}{lll}
[\Lvr r {m+\frac r \lambdam},\Gr s {n+\frac {s+\frac12}\lambdam}]
&=&(\frac12(m+\frac r\lambdam)-(n+\frac {s+\frac12}\lambdam))
\Gr{r+s}{m+n+\frac{r+s+\frac12}\lambdam}\\[1em]
[\Gr r {m+\frac {r+\frac12}\lambdam},
\Gr{s}{n+\frac {s+\frac12}\lambdam}]_+&=&
2\Lvr {r+s+1}{m+n+\frac{r+s+1}\lambdam}\\[1em]
&&+\frac{\ch}3 ((m+\frac{r+\frac12}\lambdam)^2-\frac 14)
\delta_{m+n+\frac{r+s+1}\lambdam,\,0}
.\end{array}\labl{2.9}
The $\widehat{NS}$ orbifold supercurrents also satisfy the periodicity 
relation \erf {periodicity} with $\epsilon =\frac12$.\\[2em]
\underline{\bf  $\widehat{R}$ sector}\\[2em]
In the  $\widehat{R}$ sector, the $\widehat{R}$ orbifold supercurrents 
$\Gr r {m+\frac r\lambdam}$ satisfy 
\be\begin{array}{lll}
[\Lvr r {m+\frac r \lambdam},\Gr s {n+\frac s\lambdam}]
&=&(\frac12(m+\frac r\lambdam)-(n+\frac s\lambdam))
\Gr{r+s}{m+n+\frac{r+s}\lambdam}\\[1em]
[\Gr r {m+\frac r\lambdam},
\Gr s{n+\frac s\lambdam}]_+&=&
2\Lvr {r+s}{m+n+\frac{r+s}\lambdam}
+\frac{\ch}3 ((m+\frac r\lambdam)^2-\frac 14)
\delta_{m+n+\frac{r+s}\lambdam,\,0}
\end{array}\labl{2.10}
and the periodicity condition \erf{periodicity} with $\epsilon=0$.

\subsection{Orbifold N=2 superconformal algebra}
The orbifold N=2 superconformal algebra has the counterparts 
$\widehat{NS}$, $\widehat{R}$, and $\widehat{T}$
of the usual sectors $NS$, $R$, and $T$(twisted), 
all of which are included in the uniform notation for the generators
\be \Lvr r {m+\frac r\lambdam},
\, \Gr {i,r} {m+\frac {r+\delta^i}\lambdam},
\, \Jr r {m+\frac {r+\epsilon}\lambdam}
,\,\,\,\,\,\,i=1,2,\,\,\,r=0\ldots
\lambdam-1\ee
\be \begin{array}{rl}
\widehat{NS}:&\epsilon=0,\,\,\,\delta^1=\delta^2=\frac12\\[1em]
\widehat{R}:& \epsilon=\delta^1=\delta^2=0\\[1em]
\widehat{T}:& \epsilon=\delta^2=\frac12,\,\,\,\delta^1=0
\end{array}\ee
where $i=1,2$ labels the two sets of orbifold supercurrents
and $\hat J^{(r)}$ is the set of orbifold $U(1)$ currents. In addition to
the orbifold Virasoro algebra \erf{2.5}, the N=2 system satisfies
\be \begin{array}{lll}
[\Lvr r {m+\frac r \lambdam},\Gr {i,s} {n+\frac {s+\delta^i}\lambdam}]
&=&(\frac12(m+\frac r\lambdam)-(n+\frac {s+\delta^i}\lambdam))
\Gr{i,r+s}{m+n+\frac{r+s+\delta^i}\lambdam}\\[1em]
[\Lvr r {m+\frac r \lambdam},\Jr s {n+\frac {s+\epsilon}\lambdam}]&=&
-(n+\frac{s+\epsilon}\lambdam)\Jr{r+s}{m+n+\frac{r+s+\epsilon}\lambdam}\\[1em]
[\Jr r {m+\frac {r+\epsilon}\lambdam},\Jr s {n+\frac {s+\epsilon}\lambdam}]
&=&\frac {\hat c}3
(m+\frac{r+\epsilon}\lambdam)\delta_{m+n+\frac{r+s+2\epsilon}\lambdam,\,0}\\[1em]
[\Jr r{m+\frac {r+\epsilon}\lambdam},\Gr{i,s}{n+\frac {s+\delta^i}\lambdam}]
&=& \ii\epsilon_{ij} \Gr {j,r+s+\epsilon+\delta^i-\delta^j}
{m+n+\frac {r+s+\epsilon+\delta^i}\lambdam}\\[1em]
[\Gr{i,r}{m+\frac {r+\delta^i}\lambdam},
\Gr{j,s}{n+\frac {s+\delta^j}\lambdam}]_+&=&
2\delta_{ij}\Lvr {r+s+\delta^i+\delta^j}{m+n+\frac
{r+s+\delta^i+\delta^j}\lambdam}\\[1em]
&+&\ii\epsilon_{ij}(m-n+\frac{r-s+\delta^i-\delta^j}\lambdam)
\Jr{r+s+\delta^i+\delta^j-\epsilon}{m+n+\frac{r+s+\delta^i+\delta^j}
\lambdam}\\[1em]
&+&\frac{\hat c}3 ((m+\frac{r+\delta^i}\lambdam)^2-\frac 14)\delta_{ij}
\delta_{m+n+\frac{r+s+\delta^i+\delta^j}\lambdam,\,0}
\end{array}\labl{2.13}
and the periodicity conditions \erf{periodicity}.

\subsection{Orbifold $\Wt$ algebra}
This algebra has, in addition to the orbifold Virasoro generators,
the $\hat W$ generators 
\be \Whr r {m+\frac r\lambdam},
\,\,\,r=0\ldots\lambdam-1,\,\,\,m\in\zet\ee
which satisfy the orbifold $\Wt$ algebra
\be \begin{array}{l}
[\Lvr r {m+\frac r\lambdam},\Whr s {n+\frac s\lambdam}]=
(2(m+\frac r\lambdam)-(n+\frac s\lambdam))\Whr {r+s} {m+n+\frac
{r+s}\lambdam}\\[1em]
[\Whr r {m+\frac r\lambdam},\Whr s {n+\frac s\lambdam}] =
\frac {16\lambdam}{22\lambdam+5\hat c}(m-n+\frac {r-s}\lambdam)
\lamhr {r+s}{m+n+\frac{r+s}\lambdam}\\[1em]
+(m-n+\frac {r-s} \lambdam)(\frac1{15}(m+n+\frac{r+s+2}\lambdam)
(m+n+\frac{r+s+3}\lambdam)-\frac16(m+\frac{r+2}\lambdam)
(n+\frac{s+2}\lambdam))\Lvr {r+s}{m+n+\frac{r+s}\lambdam}\\[1em]
+\frac{\hat c}{360}((m+\frac r\lambdam)^2-4)((m+\frac r\lambdam)^2-1)
(m+\frac r\lambdam) \delta_{m+n+\frac {r+s}\lambdam,\,0}\,.
\end{array}
\labl{2.15}
The composite operators $\lamhr r{m+\frac r\lambdam},\,\,\,r=
0\ldots\lambdam-1$ are constructed with the normal-ordered products
of two orbifold Virasoro operators
\be\begin{array}{lll}
\lamhr r{m+\frac r\lambdam}&=&\frac 1\lambdam
\sum_{s=0}^{\lambdam-1}\sum_n \nord \Lvr s {n+\frac s \lambdam}
\Lvr {r-s} {m-n+\frac {r-s}\lambdam} \nord\\[1em]
&&-(\frac 3{10}(m+\frac {r+2}\lambdam)(m+\frac {r+3}\lambdam)+
\frac {\hat c } {12\lambdam} (1-\frac 1 {\lambdam^2}))
\Lvr r {m+\frac r\lambdam}\\[1em]
&&+\frac {\hat c}{1440}(1-\frac 1 {\lambdam^2})(\frac{119}{\lambdam^2}
-11) \delta_{m+\frac r\lambdam,\,0}\,  \end{array}\ee
\be\begin{array}{lll}
\nord \Lvr r {m+\frac r \lambdam} \Lvr s {n+\frac s \lambdam}\nord
&\equiv&\theta(m\geq -\frac {r+1}\lambdam)\Lvr s {n+\frac s \lambdam}
\Lvr r {m+\frac r \lambdam}\\[1em]&&+
\theta(m < -\frac{r+1}\lambdam)\Lvr r {m+\frac r \lambdam}
 \Lvr s {n+\frac s \lambdam}\,
\end{array}\ee
and the periodicity of all the operators in the system is governed by
\erf{periodicity} with $\epsilon=0.$

\subsection{Orbifold induction procedure}
The orbifold algebras above and other orbifold algebras, such as the 
orbifold ${\mbox{${\rm W}_n$}}$ algebras
for $n>3$, can be obtained by an {\it orbifold induction 
procedure} from the infinite-dimensional algebras of the mother \cft.
We discuss here the mode form of the induction procedure,
deferring its local form until Section 3.

In the induction procedure, every Virasoro primary field $A_\Delta$ of
conformal weight $\Delta$ in the mother theory defines a set of $\lambdam$
orbifold fields $\hat A_\Delta^{(r)},\,\,\,r=0\ldots\lambdam-1.$
We consider here only integer and half-integer moded
primary fields with modes $A_\Delta(m+\epsilon)$
where $\epsilon=0,\frac12.$
(The induction procedure for general primary fields is discussed in 
Section 3.)
Then the modes of the orbifold fields are given by the following definition:
\be \Adhr r {m+\frac {r+\epsilon}\lambdam}\equiv\lambdam^{1-\Delta}
A_{\Delta}(\lambdam m +r+\epsilon).\labl{2.18}
The simplest application of equation \erf{2.18} is the induction of 
the orbifold affine algebra \erf{2.1} via the relation
\be
\Jahr a r {m+\frac r\lambdam}=J_a(\lambdam m+r)
\ee
from the general affine algebra \cite{KM,baha}
\be [J_a(m),J_b(n)]=\ii 
f_{ab}^{\,\,\,\,c}J_c(m+n)+k\eta_{ab}m\delta_{m+n,\,0}\ee
of the mother theory at level $k$. One finds that 
\be
\hat k =\lambdam k
\ee
where $\hat k$ is the level of the orbifold affine algebra.

For the quasi-primary fields $L$ and $\Lambda$, the induction relations read
\be \Lvr r {m+\frac r\lambdam}\equiv\frac 1\lambdam \Lv {\lambdam m+r}+
\frac {\hat c} {24}(1-\frac 1{\lambdam^2}) \,\delta_{m+\frac r\lambdam,\,0}
\labl{2.19}
\be \lamhr r {m+\frac r\lambdam}\equiv\lambdam^{-3}\Lambda(\lambdam m+r)
-\frac{\hat c(22\lambdam+5\hat c)}{2880\lambdam}(1-\frac
1{\lambda^2})^2 \delta_{m+\frac r\lambdam,\,0}
\,\ee
and one finds that the central charge of the orbifold Virasoro algebra 
\erf{2.5} is 
\be \hat c=\lambdam c \ee
where $c$ is the central charge of the Virasoro algebra in the mother 
theory. The same orbifold central charge $\hat c$
appears in the orbifold 
superconformal and $\Wt$ algebras.

Special cases of the induction relations \erf{2.18} and \erf{2.19}
are known in the literature \cite{kawa5,fukS,bouw6}.
These special cases correspond to the induction of what we will call 
integral subalgebras (see Subsection 2.7) of the orbifold affine, orbifold  
Virasoro and orbifold superconformal algebras. It was also emphasized in 
Ref. \cite{fukS} that these subalgebras of the orbifold algebras are 
operative in the twisted sectors of cyclic orbifolds. \\[2em]
\underline{\bf Unitarity}\\[2em]
The orbifold induction procedure also implies that the states of the 
twisted sector of the orbifold are the states of the mother conformal field
theory and the induced inner product of states is also the same as in the 
mother theory.

Although non-unitary conformal field theories can also be considered, we 
will generally assume
in this paper that the mother conformal field theory is a unitary conformal
field theory with a unique $\SltR$-invariant vacuum state $\mket 0$ and 
corresponding vacuum field $\phi_0=\bfe$. In the orbifold the state 
$\mket 0$ is the ground state of the twisted sector (twist field) with 
conformal weight 
\be \hat\Delta=\frac{\hat c}{24}(1-\frac1{\lambdam^2})\ee
as measured by the orbifold Virasoro generator $\Lvr 0 0$.

The induction procedure from a unitary mother conformal field theory 
gives the twisted 
sectors of the unitary cyclic orbifolds, where the adjoint operations are 
defined by \be \Lvr r{m+\frac r\lambdam}^{\dagger}=
\Lvr {\lambdam-r} {-(m+1)+\frac{\lambdam-r}\lambdam} \ee
\be \Adhr r{m+\frac {r+\epsilon}\lambdam}^{\dagger}=
\Adhr {\lambdam-2\epsilon-r} {-(m+1)+\frac{(\lambdam-2\epsilon-r)
+\epsilon}\lambdam}. \ee
These relations are nothing but the images of the adjoint operations 
in the unitary mother \cft.

\subsection{Subalgebras of the orbifold algebras}
In this subsection, we discuss the {\it integral subalgebras} 
of the orbifold algebras and two 
$\SltR$ subalgebras of the orbifold Virasoro algebras. The integral 
subalgebras of the orbifold algebras are isomorphic to the corresponding 
infinite-dimensional algebras of the mother theory.

As a related remark, it is not difficult to see for $\lambdam=N^2$ that the 
orbifold affine algebra generated by $\hat J_a^{(r)},\,\,\,r=0\ldots 
N^2-1$ contains a subalgebra (generated by $\hat J_a^{(r)}$ with $r$ 
divisible by $N$)  which is a twisted affine Lie algebra \cite{KAc3}. 
This twisted 
affine algebra can be obtained by an outer-automorphic twist of $\lambdam$
copies of a mother affine algebra at level $k$ by the $\zet_\lambdam$
permutation symmetry of the copies.\\[2em]
\underline{\bf Integral affine subalgebra}\\[2em]
In the orbifold affine algebra \erf{2.1}, the orbifold currents 
$\Jahr a 0 m,\,\,\,a=1\ldots\dim\g,\,\,\,m\in\zet$ form an integral affine
subalgebra which is an ordinary affine algebra at level $\hat k=\lambdam 
k$. This integral subalgebra 
was studied in Refs. \cite{kawa5,bouw6}, and was 
called the winding subalgebra in Ref.\cite{kawa5}.\\[2em] 
\underline{\bf Integral Virasoro subalgebra}\\[2em]
In the orbifold Virasoro algebra \erf{2.5}, the generators 
$\Lvr 0 m,\,\,\,m\in\zet\,\,$ form an integral Virasoro subalgebra (which is 
an ordinary Virasoro algebra with central charge $\hat c=\lambdam c$). 
This integral subalgebra was studied in Refs. \cite{kawa5,fukS,bouw6}.\\[2em]
\underline{\bf Integral N=1 subalgebras}\\[2em]
The integral superconformal subalgebras 
of the orbifold N=1 superconformal algebras \erf{2.9} and \erf{2.10}
are collected in the following 
table.
\be \begin{array}{cccc}
&\mbox{\underline{ Orbifold sector}}\,
&\mbox{\underline{ Generators}}\,
&\mbox{\underline{Integral N=1 subalgebra}}\\[1em]
\lambdam=2l+1&\widehat{NS}&\hat G^{(l)},\hat L^{(0)}\,&NS\\[1em]
             &\widehat{R} &\hat G^{(0)},\hat L^{(0)}\,&R \\[1em]
\lambdam=2l  &\widehat{NS}&-&-\\[1em]
	     &\widehat{R} &\hat G^{(l)},\hat L^{(0)}\,&NS\\[1em]
             &\widehat{R} &\hat G^{(0)},\hat L^{(0)}\,&R \\[1em]
.\end{array}\ee
This table shows in particular that, for even $\lambdam$, the $\widehat{R}$ 
orbifold algebra contains both an ordinary NS and an ordinary R 
superconformal 
subalgebra, while the $\widehat{NS}$ orbifold algebra contains no 
superconformal subalgebras. 
\\[2em]
\underline{\bf Integral N=2 subalgebras}\\[2em]
The integral N=2  superconformal subalgebras of the  orbifold
$N=2$ superconformal algebras \erf{2.13} are given in the table below.
\be \begin{array}{cccc}
&\mbox{\underline{ Orbifold sector}}\,
&\mbox{\underline{ Generators}}\,
&\mbox{\underline{Integral N=2 subalgebra}}\\[1em]
\lambdam=2l+1  &\widehat{NS}&\hat G^{(i,l)},\hat J^{(0)},\hat 
L^{(0)}\,&NS\\[1em]
               &\widehat{R} &\hat G^{(i,0)},\hat J^{(0)},\hat L^{(0)}\,&R 
\\[1em]
               &\widehat{T} &\hat G^{(1,0)},\hat G^{(2,l)},
			\hat J^{(l)},\hat L^{(0)}\,&T \\[1em] 
\lambdam=2l    &\widehat{NS}&-&-\\[1em]
	       &\widehat{R} &\hat G^{(i,l)},\hat J^{(0)},\hat L^{(0)}\,&NS\\[1em]
               &\widehat{R} &\hat G^{(i,0)},\hat J^{(0)},\hat L^{(0)}\,&R 
\\[1em]
               &\widehat{R} &\hat G^{(1,0)},\hat G^{(2,l)},
			\hat J^{(l)},\hat L^{(0)}\,&T \\[1em] 
	       &\widehat{T} &-&- 
\end{array}\ee
Except for the $\widehat T$ sector, these 
integral subalgebras have been studied in Ref. \cite{fukS}, where it was 
also noted for even $\lambdam$ that the $\widehat{NS}$ sector of the 
orbifold is eliminated in string theory by the GSO projection \cite{glso}.
\\[2em]
\underline{\bf $\Wt$ integral subalgebras}\\[2em]
The orbifold $\Wt$ algebra \erf{2.15} has no integral $\Wt$ subalgebras, so 
orbifoldization has, in this case, removed an infinite-dimensional 
symmetry of the mother theory.\\[2em]
\underline{\bf $SL(2,{\sf I}\!{\sf R})$ subalgebras}\\[2em]
Any integral Virasoro subalgebra contains an $\SltR$ subalgebra whose 
generators are 
\be \Lvr 0 1, \,\Lvr 0 {-1}, \,\Lvr 0 0.\ee
In addition, each
orbifold Virasoro algebra contains a centrally-extended $\SltR$ subalgebra
\be \begin{array}{lll}
[\Lvr 1 {\frac 1 \lambdam},\Lvr {\lambdam-1} {-1+\frac {\lambdam -1}
\lambdam}]
&=&\frac 2 \lambdam \Lvr 0 0 -\frac {\hat c} {12\lambdam} (1-\frac 1
{\lambdam^2})\\[1em]
[\Lvr 0 0,\Lvr 1 {\frac 1 \lambdam}]&=&-\frac 1 \lambdam \Lvr 1 {\frac 1
\lambdam} \\[1em]
[\Lvr 0 0,\Lvr {\lambdam-1} {-1+\frac {\lambdam -1} \lambdam}] &=&
\frac 1 \lambdam \Lvr {\lambdam-1} {-1+\frac {\lambdam -1}
\lambdam} \end{array}\labl{sltr}
which is the image of the $\SltR$ algebra of the mother theory.
The ground state $\mket 0$ of the twisted sector of the orbifold is not 
$\SltR$-invariant under either of these $\SltR$ subalgebras, and we note in
particular that 
\be \begin{array}{c}
\Lvr 1 {\frac 1 \lambdam} \mket 0 = \Lvr {\lambdam-1}
{-1+\frac{\lambdam-1}\lambdam}\mket 0 =0\\[1em]
\Lvr 0 0 \mket 0 = \frac \ch {24} (1-\frac 1 {\lambdam^2})\mket 0 \, 
\end{array}\labl{2.32}
for the centrally-extended $\SltR$. Curiously however, we will see in 
Subsection 3.6 that the twisted sector of the orbifold contains
$\SltR$ Ward identities associated to the centrally-extended $\SltR$.


\sect{Twisted sectors on the sphere}
\subsection{Principal primary fields}
Any Virasoro primary state $\mket \Delta$ in the mother theory is 
proportional to the state $\mket{\hat\Delta}$ in the orbifold, which is
primary with respect to the integral Virasoro subalgebra
\be \Lvr 0 {m\geq 0}\mket{\hat \Delta}=\delta_{m,\,0}\hat\Delta
\mket{\hat \Delta}.\labl{3.1}
The conformal weight $\hat\Delta$ in \erf{3.1} is
\be \hat \Delta=\frac \Delta\lambdam +\frac{\hat c}
{24}(1-\frac1{\lambdam^2}). \ee
There is also an infinite number of other primary states of the integral
Virasoro subalgebra, which are not Virasoro primary in the mother theory.
At induction order $\lambdam$, examples of such states include all 
the descendants in the $\Delta$ module whose level in the module is less 
than or 
equal to $\lambdam-1$.
We focus in particular on the multiplet of $\lambdam$ states
\be 
\begin{array}{c}
\left(L(-1)\right)^{\lambdam-1-r}\mket \Delta\sim
\left(\Lvr {\lambdam-1} {-1+\frac{\lambdam-1}\lambdam}\right)^{\lambdam-1-r}
\mket
{\hat\Delta},\,\,\,\,\,\,r=0\ldots\lambdam-1 \\[1em]
\hat\Delta_r=\frac{\lambdam-1-r+\Delta}\lambdam+\frac{\hat
c}{24}(1-\frac1{\lambdam^2})\end{array}
\labl{states}
which, up to normalization, we will call the {\it principal primary states}
of the twisted sector. All these states 
are Virasoro primary
with conformal weights $\hat \Delta_r$
under the integral Virasoro subalgebra, including the state
$\mket{\hat\Delta}$ with 
$\hat\Delta=\hat\Delta_{\lambdam-1}$.

The principal primary states have the distinction that all other 
primary states under the integral Virasoro subalgebra have conformal
weights $\hat h=\{\hat\Delta_r\}+n$ which differ by a non-negative integer 
$n$
from the conformal weights $\hat \Delta_r$ in \erf{states}.
In what follows, we construct the principal primary
fields $\phidhhr r (z),\,\,\,r=0\ldots\lambdam-1$ of the twisted 
sector of the orbifold, which, as we shall see, are interpolating fields,
acting 
\futnote{As a consequence, the principal primary fields do not include
the so-called twist field of the orbifold - which acts between states
in the twisted and untwisted sectors.}
between states in the twisted sector, for the principal primary states 
\erf{states}. 

We begin with a primary field $\varphi_\Delta$ of conformal weight $\Delta$
in the mother theory, which satisfies
\be [\Lv m,\phid(z)]=z^m(z\delz+\Delta(m+1))\phid(z). \labl{prop-ph}
We will assume that the primary field has a diagonal monodromy such that
\be\begin{array}{c}
\phid(z\mono) = \phid(z) \eE^{2\pi\ii\theta} \\[1em]
\phid(z) [ \Lv m, \theta] =0 \\[1em]
\theta \mket 0 = 0
\end{array}
\ee
examples of which are provided by the familiar abelian vertex operators
and by all simple currents \cite{bern,fuge,halp7,scya,intr}.

The {\it principal primary fields} $\,\,\phidhhr r 
(z),\,\,\,r=0\ldots\lambdam-1$
of the twisted sector of the orbifold are defined as
\be \begin{array}{lll}
\phidhhr r(z) &\equiv& \sum_{s=0}^{\lambdam-1} \phidhh(z\eE^{2\pi\ii s})
\eE^{\frac{2\pi\ii s}\lambdam (-\theta +r +1 +\Delta(\lambdam-1))} \\[1pt]
&=& \rho(z)^\Delta \sum_{s=0}^{\lambdam-1} \phid(z^{\frac1\lambdam}
\eE^{\frac{2\pi\ii s}\lambdam}) \eE^{\frac{2\pi\ii s}\lambdam (-\theta +r
+1 )} \end{array}\labl{de1}
where 
\be
\rho(z) \equiv \frac 1\lambdam z^{{\frac 1 \lambdam} -1}
\labl{3.7}
\be
\phidhh(z) \equiv \rho(z)^\Delta \phid(z^{\frac1\lambdam})
\labl{3.8}
\be
[\Lvr r {m+\frac r \lambdam}, \phidhh(z)] = z^{m+\frac r \lambdam}
(z \delz + \Delta(m+\frac r \lambdam+1)) \phidhh(z)  .
\labl{3.*}
The commutation relation \erf{3.*} follows from 
\erf{prop-ph}, \erf{3.7} and \erf{3.8}. 

The principal primary fields have the following properties,
\be \phidhhr{r\pm\lambdam} (z)= \phidhhr r (z)\,  \ee
\be
 \phidhhr r(z\mono)  =  \phidhhr r(z)
\eE^{\frac{2\pi\ii }\lambdam (\theta -r -1 -\Delta(\lambdam-1))}
\labl{3.10}
\be
\sum_{r=0}^{\lambdam-1} \phidhhr r(z) = \lambdam \phidhh(z)
\labl{split}
\be [\Lvr r {m+\frac r \lambdam}, \phidhhr s(z)] = z^{m+\frac r \lambdam}
(z \delz + \Delta(m+\frac r \lambdam+1)) \phidhhr {s+r} (z) \, \labl2
which follow from eqs. (3.4-9).

The reader should bear in mind that the field $\phidhh(z)$ in \erf{3.8}
is a locally-conformal transformation, 
by $z\to z^{\frac1\lambdam}$, 
of the Virasoro primary field $\varphi_{\Delta}(z)$,
 where 
the quantity $\rho(z)$ in \erf{3.7}
\be \rho(z)=\frac{\partial (z^{\frac1\lambdam})}{\partial z} \ee
is the Jacobian of the transformation. The field
$\phidhh(z)$ therefore lives on the root coverings of the sphere.
Moreover, equations \erf{de1} and \erf{split} show that the principal 
primary fields 
are the fields with definite monodromy which result from the 
monodromy decomposition of the conformally-transformed field $\phidhh(z).$
This conformal transformation has appeared in free-field examples in
Ref. \cite{hava}.

The principal primary fields $\phidhhr r$ create the principal primary 
states $\,\,\mket{\hat\Delta_r},\,\,\,r=0\ldots\lambdam-1$  as follows, 
\be \begin{array}{ll}
\mket{\hat\Delta_r} &\equiv \lim_{z\to 0}
z^{\Delta(1-\frac 1\lambdam)-1+\frac{r+1}\lambdam} \phidhhr r (z) \mket 0
\\[1em]&
= \lambdam^{1-\Delta} \frac{(\partial^{\lambdam-r-1} 
\phid)(0)}{(\lambdam-r-1)!} \mket 0 
= \frac{\lambdam^{1-\Delta} L(-1)^{\lambdam-r-1}}
{(\lambdam-r-1)!} \mket \Delta \\[1em]&
= \frac{\lambdam^{\lambdam-r-1} 
\{\Lvr{\lambdam-1}{-1+\frac{\lambdam-1}\lambdam}\}^{\lambdam-r-1}}
{(\lambdam-r-1)!} \mket {\hat\Delta_{\lambdam-1}}  \\[1em]
\mket{\hat\Delta_{\lambdam-1}} &= \lambdam^{1-\Delta} \lim_{z\to0}\phid(z) 
\mket 0 = \lambdam^{1-\Delta} \mket{\Delta}.
\end{array}\labl{3.15}
The relations in \erf{3.15} identify the principal primary states as those
shown in eq. \erf{states}.
Using these results, the following algebraic properties of the principal 
primary states $\mket{\hat\Delta_r}$ are easily verified:
\be\begin{array}{c}
\Lvr r {m+\frac r\lambdam}\mket {\hat{\Delta}_s} =0,\,\,\, m>0
\\[1em]
\Lvr 0 0 \mket{\hat \Delta_r} = \hat\Delta_r \mket{\hat\Delta_{r}}
\\[1em]
\hat\Delta_r = 
(\frac{\lambdam-r-1+\Delta}\lambdam)+
\frac {\hat c}{24}(1-\frac1{\lambdam^2})
\\[1em]
\Lvr r {\frac r\lambdam}\mket {\hat\Delta_s}=0,\,\,\,
r+s\geq\lambdam
\\[1em]
\Lvr r {\frac r\lambdam} \mket{\hat \Delta_s} =
(\frac{\Delta(r+1)+\lambdam-s-r-1}\lambdam \, )
\mket{\hat\Delta_{s+r}},\,\,\,\,r=1\ldots\lambdam-1
\\[1em]
\Lvr {\lambdam-1} {-1+\frac{\lambdam-1}\lambdam}
\mket{\hat\Delta_{r}}
= (1-\frac r\lambdam) \mket{\hat\Delta_{r-1}} \, .
\end{array}\labl{3.16new}
The first three relations in \erf{3.16new} include the fact that the 
principal 
primary states are primary under the integral Virasoro subalgebra, with 
conformal weights $\hat\Delta_r$, as they should be. The state 
$\Lvr {\lambdam-1} {-1+\frac{\lambdam-1}\lambdam}\mket{\hat\Delta_{0}}$
is not in general Virasoro primary under the integral Virasoro subalgebra.
\\[2em]
\underline{\bf Example: ground state of the twisted sector}\\[2em]
Some of the principal primary fields and states may be 
null. This is seen in the simplest example, which is the ground state
of the twisted sector. The vacuum field of the mother theory is 
$\varphi_0(z) = \bfe$, so that we obtain for the corresponding
set of principal primary fields,
\be\begin{array}{c}
\phivhhr r(z) =
\lambdam\delta_{r+1,\,0\bmod\lambdam}=
\lambdam \delta_{r,\lambdam-1}
\\[1em]
\lim_{z\to0}\phivhmr{\lambda-1} z \mket 0 = \lambdam \mket 0 .
\end{array} \labl{3.13}
These fields and states are null for $r\neq\lambdam-1$ because the state 
$\Lvr {\lambdam-1} {-1+\frac{\lambdam-1}\lambdam}\mket 0$ is null,
which corresponds to the fact that the ground state of this twisted sector
is non-degenerate.
\\[2em]
\underline{\bf Modes}\\[2em]
For those Virasoro primary fields of the mother theory which have 
integer or half-integer conformal weight $\Delta$ and trivial monodromy,
the discussion above can be expressed in modes. In the mother theory, the 
mode resolution of $\phid(z)$ can be written as
\be\begin{array}{c}
\phid(z)  = \dsum_{m\in\zet} \phidm {m+1-\Delta} z^{-(m+1-\Delta)}  
z^{-\Delta} \\[1em]
 \lim_{z\to0}\phid(z) \mket 0  = \phidm {-\Delta} \mket 0
\\[1em]
\phidm {-\Delta+m} \mket 0 = 0,\,\,\,\,\,m>0 .
\end{array} \ee
Then one obtains the mode form of the principal primary fields and
states,
\be
\phidhhr r(z)= \sum_{m\in\zet} \phidhmr r{m+\frac{r+1-\Delta}\lambdam}
z^{-(m+\frac{r+1-\Delta}\lambdam)} z^{-\Delta}
\ee
\be
\phidhmr r {m+\frac{r+1-\Delta}\lambdam} =
\lambdam^{1-\Delta} \phidm{\lambdam m+r+1-\Delta}
\labl{3.16}
\be
[\Lvr r{m+\frac r\lambdam}, \phidhmr s{n+\frac{s+1-\Delta}\lambdam}]
= ((m+\frac r\lambdam)(\Delta-1) - (n+\frac{s+1-\Delta}\lambdam))
\phidhmr{s+r}{n+m+\frac{s+r+1-\Delta}\lambdam}
\ee
\be
[\Lvr 0 0,\phidhhr r (-1+\frac{r+1-\Delta}\lambdam)]=
(\frac {\lambdam-r-1+\Delta}\lambdam)
\phidhhr r(-1+\frac{r+1-\Delta}\lambdam)
\labl{3.23}
\be
\mket{\hat{\Delta}_r} =\, \phidhmr r {-1+\frac{r+1-\Delta}\lambdam} \mket 0 
= \lambdam^{1-\Delta} \phidm{-\Delta+r+1-\lambdam} \mket 0 .
\ee
The coefficient on the right side of eq. \erf{3.23} is the non-shift
part of the conformal weight $\hat\Delta_r$ in \erf{3.16new}.
We also note that (except for the orbifold Virasoro generators) the 
generators of
the orbifold algebras are the modes of principal primary fields of the 
orbifolds, although the mode-labeling convention \erf{3.16} of 
the principal primary fields can differ by a cyclic relabeling from the 
mode-labeling convention \erf{2.18} of the orbifold algebras.

\subsection{OPE's and correlators of principal primary fields}

In this subsection, we will use the orbifold induction procedure to 
compute the operator product expansion (OPE) of two principal primary fields.

We begin with the OPE of two Virasoro primary fields in the mother theory,
\be \phidn 1(z) \, \, \phidn 2(w) 
= \sum_i \frac{F_{12}^{\,\,\,\,\,i}}{(z-w)^{\Delta_1+\Delta_2-\Delta_i}} \, 
( 1 +\sum_{p=1}^\infty a_p^i (z-w)^p \partial_w^p ) \chidn i(w) \, 
\labl{555}
where $\chidn i$ are Virasoro quasi-primary fields. The constants $a_p^i$ 
in \erf{555} are fixed by $\SltR$ invariance 
\be a_p^i=\frac1{p!}\prod_{q=0}^{p-1}\frac{(\Delta_i+\Delta_1-\Delta_2+q)}
{(2\Delta_i+q)} \ee
and further relations between $F$'s follow from
 conformal invariance when sets of quasi-primary fields are in the same 
Virasoro module. 

Because OPE's are form invariant under conformal transformations \cite{bepz}
and $\phidhh$ is a conformal transformation of $\phid$ (see Subsection 3.1),
it follows that 
\be \phidohh(z) \, \, \phidthh(w)
= \sum_i \frac{F_{12}^{\,\,\,\,\,i}}{(z-w)^{\Delta_1+\Delta_2-\Delta_i}} \,
( 1 +\sum_{p=1}^\infty a_p^i (z-w)^p \partial_w^p ) \chidihh(w) \, 
\labl{3.18}
where $\chidihh(w)$ is the corresponding conformal transformation
(by $z\to z^{\frac1\lambdam}$) of $\chidn i$. 

For simplicity, we now assume that the mother field $\phidn 1$ has trivial 
monodromy, which implies that
\be \theta_i=\theta_2,\,\,\,\,\,\Delta_1+\Delta_2 
-\Delta_i \in \zet\,.\ee
where $\theta_i$ is the monodromy of $\chidihh$.
Then one computes from \erf{de1} and \erf{3.18} 
the OPE's of two principal primary fields,
\be \phidohhr r(z) \, \, \phidthhr s(w)
= \sum_i \frac{F_{12}^{\,\,\,\,\,i}}{(z-w)^{\Delta_1+\Delta_2-\Delta_i}} \,
( 1 +\sum_{p=1}^\infty a_p^i (z-w)^p \partial_w^p )
\chidihhr {r+s+1+\Delta_i-\Delta_1-\Delta_2}(w)+\reg .  
\labl{3.20}
where
\be\begin{array}{c}
 \chidihhr r(w)=\sum_{s=0}^{\lambdam-1}\eE^{\frac {2\pi\ii s}\lambdam
(r+1-\theta_i+\Delta_i(\lambdam-1))}\chidihh (w\eE^{2\pi\ii s})\\[1em]
\chidihhr {r+\lambdam}(w)=\chidihhr r(w)
\end{array}\ee
is obtained for the monodromy decomposition of the conformal 
transformation of the quasi-primary fields. Given the monodromy algebra of 
$\theta_1\neq0$ with $\phidn 2$ in the mother theory, these OPE's can be 
obtained in full generality.

Using the monodromy decomposition \erf{de1}, we may also express the 
correlators of any number of  principal primary fields in terms of vacuum 
averages in the mother theory,
\be \begin{array}{l}
\bra 0 \phidnhhr 1 {r_1} (z_1)...\phidnhhr n {r_n} 
(z_n)\mket 0 = (\prod_{i=1}^n 
\rho^{\Delta_i}(z_i))\sum_{s_1=0}^{\lambdam-1}...\sum_{s_n=0}^{\lambdam-1}
\eE^{\frac {2\pi\ii} \lambdam\sum_{j=1}^{\lambdam-1}s_j(r_j+1)}
\Phi^{(s)}_{(\Delta)}(z)
\\[2em]
\Phi^{(s)}_{(\Delta)}(z)=
\langle \phidn 1 (z_1^{\frac 1\lambdam}\eE^{\frac {2\pi\ii s_1}\lambdam})
\eE^{-\frac {2\pi\ii s_1\theta_1}\lambdam}...
\phidn {n-1} (z_{n-1}^{\frac 1\lambdam}\eE^{\frac {2\pi\ii 
s_{n-1}}\lambdam})   \eE^{-\frac {2\pi\ii s_{n-1}\theta_{n-1}}\lambdam}
\phidn n (z_n^{\frac 1\lambdam}\eE^{\frac {2\pi\ii s_n}\lambdam})    
\rangle \,.
\end{array} \labl{3.22}
Although the monodromy algebra of $\theta$'s and $\varphi$'s is needed to 
evaluate these quantities in the general case, the evaluation is 
straightforward for fields in the mother theory with integer and half-integer
conformal weight $\Delta$ and trivial monodromy. As an example, we 
consider the two-point correlators, which satisfy
\be
\langle \phid(z)\phid(w) \rangle=\frac B{(z-w)^{2\Delta}}
\ee
in the mother theory. Then eq. \erf{3.22} gives
\be\begin{array}{c}
\bra 0 \phidhhr r (z) \phidhhr s (w) \mket 0 =
\frac {\lambdam B}{(2\Delta-1)!}(\frac wz)^{\frac{\Delta(\lambdam+1)
-(s+1)}\lambdam} (w^{1-2\Delta}\delta_{r+s+2-2\Delta,\,0\bmod\lambdam})
D_{\Delta}(w)\frac1{z-w}\\[2em]
D_{\Delta}(w) =
\prod_{k=0}^{2\Delta-2}(w\partial_w+\frac{\lambdam-s+k}\lambdam)
\end{array}\labl{<ff>}
for the two-point correlators in the twisted sector of the orbifold.

\subsection{Local form of the orbifold stress tensor}

The stress tensor of the mother theory is moded as
\be T(z) \equiv \sum_m \Lv m  z^{-m-2} \, . \ee
We define the local form of the twist classes of the orbifold stress 
tensor as
\be 
\Thhr r (z) \equiv \sum_{m\in\zet} \Lvr r{m+\frac r\lambdam} 
z^{-(m+\frac r\lambdam) -2},\,\,\,r=0\ldots\lambdam-1
\ee
\be
\Thh(z)\equiv \frac1\lambdam\sum_{r=0}^{\lambdam-1}\Thhr r(z)
\labl{3.3.*}
where the modes $\Lvr r {m+\frac r\lambdam}$ are given in \erf{2.19},
and the relation \erf{3.3.*} parallels the relation \erf{split} for 
the principal primary fields. It follows that 
\be \Thh(z) = \rho(z)^2 T(z^{\frac1\lambdam}) + \frac 
{\hat c}{24\lambdam z^2}(1-\frac1{\lambdam^2})  \labl{3.3.**}
where $\hat c =\lambdam c$ is the central charge of the orbifold Virasoro
algebra \erf{2.5}.
As expected, $\Thh(z)$ in \erf{3.3.**}
is a conformal transformation of $T(z)$ by the same
base-space transformation ($z\to z^{\frac1\lambdam}$) that gave $\phidhh(z)$
for the principal primary fields (see Subsection 3.1). The shift term in 
\erf{3.3.**} is the
Schwarzian derivative for this transformation, which can appear because
$T(z)$ is only quasi-primary under itself. This confirms the general 
prescription for the induction of quasi-primary fields seen on the right of 
eqs. \erf{3.18} and \erf{3.20}.

The twist classes $\Thhr r$ can also be viewed as the fields with definite
monodromy which result from the monodromy decomposition of $\hat T$,
\be\begin{array}{lll}
\Thhr r(z) &=& \sum_{s=0}^{\lambdam-1} \eE^{\frac{2\pi\ii s}\lambdam r}
\Thh(z\eE^{2\pi\ii s}) \\[1em]
&=&
 \rho(z)^2 \sum_{s=0}^{\lambdam-1} \eE^{\frac{2\pi\ii s}\lambdam (r+2)}
T(z^{\frac1\lambdam}\eE^{\frac{2\pi\ii s}\lambdam}) 
+ \frac{\hat c}{24z^2}(1-\frac1{\lambdam^2})\delta_{r,\,0} 
\end{array}\labl{545}
\be
\Thhr r (z\eE^{2\pi\ii})=\Thhr r (z) \eE^{-\frac{2\pi\ii r}\lambdam}
\labl{3.38new}
\be
\Thhr {r\pm\lambdam}(z)=\Thhr r(z) 
\ee
and the monodromy  relations \erf{3.38new} identify the field
\be\hat T^{(0)}(z)=\sum_{m\in\zet} \Lvr 0 m z^{-m-2}\ee
 as the chiral stress tensor of the
twisted sector of the orbifold.

Moreover, the OPE's among $\Thhr r$ and the principal primary fields 
$\phidhhr s$ may 
be computed from their monodromy decompositions and their OPE's in the mother
theory. The results are
\be \Thhr r(z) \phidhhr s(w) = (\frac \Delta {(z-w)^2}+
\frac {\partial_w} {(z-w)})\phidhhr {s+r}(w) + \reg.\labl{3.26}
\be \Thhr r(z) \Thhr s(w)=\frac{(\hat c/2)\delta_{r+s,\,0\bmod\lambdam}}
{(z-w)^4}+
(\frac 2{(z-w)^2}+ \frac {\partial_w} {(z-w)})\Thhr {r+s}(w) +\reg. 
\labl{3.27}
where eq. \erf{3.27} is the local form of the orbifold Virasoro algebra
\erf{2.5}. In obtaining \erf{3.27}, one recognizes that the coefficient 
of the central term is 
\be
\frac c2 \phivhhr {r+s-1} (z)= \frac{\hat c}2 \delta_{r+s,\,0\bmod\lambdam}
\ee
where $\phivhhr r$ is the ground state field \erf{3.13} of the twisted 
sector.

\subsection{Orbifold currents and applications} 
In this subsection, we study the orbifold currents $\Jahhr a r,\,\,\,
r=0\ldots\lambdam-1$, which satisfy the orbifold affine algebra 
\erf{2.1} at level $\hat k =\lambdam k$, and discuss various applications 
of these currents.

The local form of the orbifold currents is
\be 
\Jahhr a r (z) =\sum_{m\in\zet} \Jahr a r {m+\frac r\lambdam}
z^{-(m+\frac r\lambdam)-1}
=\sum_{s=0}^{\lambdam-1}\eE^{\frac{2\pi\ii}\lambdam rs}
\hat J_a(z\eE^{2\pi\ii s})
\labl{3.43new}
\be
\hat J_a(z)=\rho(z)J_a(z^{\frac1\lambdam})
\ee
\be \Jahr a r {m+\frac r\lambdam}\mket 0 =0,\,\,\,\,\, m\geq 0\ee 
\be \mket {\hat\Delta_r}_a=
\lim_{z\to 0}z^{\frac r\lambdam}\Jahhr a r (z)\mket 0
=\Jahr a r{-1+\frac r\lambdam}\mket 0 
= J_a(-\lambdam+r)\mket 0 \labl{44}
\be \Jahhr a r (z) \Jahhr b s (w) =
\frac{\hat k \eta_{ab}\delta_{r+s,\,0\bmod\lambdam}}{(z-w)^2}
+\frac {\ii f_{ab}^{\,\,\,\,c}\Jahhr c {r+s}(w)}{(z-w)} +\reg.\,  \labl{3.40}
\be
\bra 0 \Jahr a r z \Jahr b s w \mket 0=
(\hat k \eta_{ab} \delta_{r+s,\,0\bmod\lambdam})\,
(\frac wz)^{1-\frac s\lambdam}
\left(\frac1{(z-w)^2}+\frac{\lambdam-s}{\lambdam w}
\frac1{z-w}\right) 
\labl{<jj>}
where \erf{3.40} is the local form of the orbifold affine algebra 
\erf{2.1} and \erf{<jj>} is a special case of \erf{<ff>}.
\\[2em]
\underline{\bf Affine-Sugawara construction on the orbifold}\\[2em]
We consider next the cyclic orbifolds formed from tensor products of 
affine-Sugawara
constructions \cite{baha,halp,dafr,knza} on simple affine algebras.
Beginning with a general affine-Sugawara construction 
at level $k$ of $\g$ in the mother
theory, the induction procedure gives the orbifold affine-Sugawara
construction
\be \begin{array}{lll} 
\hat L_\g^{(r)}(m+\frac r\lambdam) &=& \frac 1\lambdam
L_\g^{ab}(k)
\sum_{n\in\zet}
\sum_{s=0}^{\lambdam-1} \nord \Jahr a s {n+\frac s\lambdam} 
\Jahr b {r-s} {m-n+\frac{r-s}\lambdam} \nord\\[1em]&& 
+\frac {{\hat c}_\g}{24}(1-\frac1{\lambdam^2})\delta_{m+\frac 
r\lambdam,\,0},\,\,\,\,\,\,\,\,r=0\ldots\lambdam-1
\end{array}\labl{3.46}
\be 
\frac1\lambdam L_\g^{ab}(k)=\frac 1\lambdam\frac {\eta^{ab}}{2k+Q_\g}
= \frac{\eta^{ab}}{2(\lambdam k)+(\lambdam Q_\g)},\,\,\,\,\,\,
{\hat c}_\g=\lambdam c_\g(k)=\frac{2\lambdam k\dim\g}{2k+Q_\g}
\labl{3.51new} \\[1pt]
\be\begin{array}{lll}
\nord \Jahr a r {m+\frac r\lambdam}\Jahr b s {n+\frac s\lambdam}\nord
&\equiv&
\theta(m+\frac r\lambdam\geq 0)\Jahr b s {n+\frac s\lambdam}
 \Jahr a r{m+\frac r\lambdam}\\[1em] &&
+\theta(m+\frac r\lambdam < 0)\Jahr a r{m+\frac r\lambdam} 
\Jahr b s {n+\frac s\lambdam}\, \end{array}\labl{3.36} 
\be
\hat L_\g^{(r)}(m+\frac r\lambdam)\mket 0 =
\delta_{m+\frac r\lambdam,\,0}\,
\frac {\hat c_\g}{24}(1-\frac 1{\lambdam^2})\mket 0 ,\,\,\,\,\,m\geq 0
\ee
in the twisted sectors of the corresponding cyclic orbifold,
where $\eta^{ab}$ is the inverse Killing metric of $\g$.
The orbifold currents in (3.50-3.53) have level $\hat k =\lambda k$.

The orbifold currents are principal primary fields under 
the orbifold affine-Sugawara construction, and we obtain
the relations
\be [\hat L_\g^{(r)}(m+\frac r\lambdam),\Jahr a s {n+\frac s\lambdam}]=
-(n+\frac s\lambdam)\Jahr a {s+r} {n+m+\frac {s+r}\lambdam}\ee 
\be [\hat L_\g^{(r)}(m+\frac r\lambdam),\Jahhr a s (z)]= 
\delz(z^{m+\frac r\lambdam+1}\Jahhr a {s+r}(z)) \ee 
\be {\hat T}_\g^{(r)} (z) \Jahhr a s (w)=(\frac 1{(z-w)^2}+
\frac 1{(z-w)}\partial_w) \Jahhr a {s+r} (w)+\reg. \labl{3.41}
which express in the orbifold the fact that the currents have conformal 
weight $\Delta=1$ in the mother theory.

One also finds that the 
principal primary states \erf{44} of the orbifold currents have conformal 
weights 
\be
\hat \Delta_r^\g=(1-\frac r\lambdam)+\frac{\hat c_\g}{24}(1-\frac
1{\lambdam^2}),~\,\,\,r=0\ldots\lambdam-1
\ee
under the integral Virasoro subalgebra of the orbifold affine-Sugawara 
construction \erf{3.46}.
Similarly, the principal primary states associated to the affine primary
states of the mother theory have orbifold conformal weights 
\be
\hat \Delta_r^\g(T)=\frac{\lambdam-r-1+\Delta_\g(T)}\lambdam\,+\,
\frac{\hat c_\g}{24}(1-\frac1{\lambdam^2}),\,\,\,\,\,\,r=0\ldots\lambdam-1
\ee
where $\Delta_\g(T)$ is the conformal weight of irrep $T$ in the mother
theory.\\[2em]
\underline{\bf Higher-level vertex operator constructions}\\[2em]
The vertex operator constructions \cite{halp3,halp6,bahn,lewi,frka,sega}
of the affine algebras are well known at level one: In the mother theory, one 
has for level one of the untwisted simply-laced algebras,
\be\begin{array}{c}
E_\alpha(z)=c_\alpha\nord \eE^{\ii\alpha\cdot Q}\nord\\[1em]
J_A(z)=\ii\delz Q_A(z)
\end{array}\labl{3.4.*}
where $\alpha\in\Delta$ with $\alpha^2=2$, $A=1\ldots{\rm rank}\, \g$
and  $c_\alpha$ is the Klein transformation. 
The string coordinate $Q$ in 
\erf{3.4.*}
involves the familiar zero-mode operators $q$ and $p$, which satisfy 
canonical commutation relations.

The orbifold induction relation \erf{2.18} then gives the vertex operator 
construction of the orbifold affine algebra at level $\hat k=\lambdam$,
\be \begin{array}{lll}
\Eahr \alpha r {m+\frac r\lambdam}&=&E_\alpha(\lambdam m+r)=
c_\alpha\oint \frac {dz}{2\pi\ii} z^{\lambdam m+r}
\nord \eE^{\ii\alpha\cdot Q(z)}\nord    \\[1em]&=&
c_\alpha\oint_{\lambdam} \frac {dw}{2\pi\ii} w^{m+\frac r\lambdam}
\rho(w)\nord \eE^{\ii\alpha\cdot \hat Q(w)}\nord\, ,
\,\,\,\,\,\,r=0\ldots\lambdam-1
\end{array}\labl{3.43}
\be
\Eahr \alpha r w =
c_\alpha \sum_{s=0}^{\lambdam-1}\eE^{\frac {2\pi\ii sr}
\lambdam} \rho(w\eE^{2\pi\ii s})
\nord \eE^{\ii\alpha\cdot \hat Q(w\eE^{2\pi\ii s})}\nord
\labl{3.43*}
\be \Jahhr A r (w)=\sum_{s=0}^{\lambdam-1}\eE^{\frac{2\pi\ii}\lambdam
rs}\hat J_A(w\eE^{2\pi\ii s})
\labl{3.43**}
where $\phidhh(z)=\rho^\Delta(z)\phid(z^{\frac 1\lambdam})$ 
and $\Delta=0$ for $Q$. In \erf{3.43}, the symbol $\oint_\lambdam$
means that the contour goes counterclockwise around the origin $\lambdam$
times. The normal ordering in \erf{3.43} and \erf{3.43*} is 
\be
\nord \eE^{\ii\alpha\cdot \hat Q(w)}\nord \,\equiv\,
\eE^{\ii\alpha\cdot q}w^{\frac{\alpha\cdot p}\lambdam}
\eE^{\ii\alpha\cdot \hat Q^-(w)}\eE^{\ii\alpha\cdot \hat Q^+(w)}
\ee
\be
\hat Q_A^\pm(w)\equiv \pm{\frac \ii\lambdam} \left\{
\sum_{r=1}^{\lambdam-1}
\frac \lambdam r\Jahr A {\pm r} {\pm\frac r\lambdam} w^{\mp\frac r\lambdam}
+\sum_{r=0}^{\lambdam-1}\sum_{m=1}^{\infty} (m+\frac r \lambdam)^{-1}
\Jahr A {\pm r} {\pm(m+\frac r\lambdam)} w^{\mp(m+\frac r\lambdam)}
\right\}\,
\labl{3.45}
and the orbifold Cartan currents $\Jahhr A {-r}$ in \erf{3.45} are defined as
\be \Jahr A {-r} {m-\frac r\lambdam}= 
\Jahr A {\lambdam-r} {m-1+\frac {\lambdam-r}\lambdam},\,\,\,\,
A=1\ldots{\rm rank}\, \g \ee
in agreement with the general relation \erf{-r}.

According to the induction procedure, the construction 
(3.61-62) for $\Jahhr a r=(\hat E_\alpha^{(r)},
\Jahhr A r)$ satisfies the orbifold affine algebra \erf{3.40}
at level $\hat k=\lambdam$. This can also be checked directly from the
OPE's
\be \nord \eE^{\ii\alpha\cdot\hat Q(z)}\nord\nord\eE^{\ii\beta\cdot\hat 
Q(w)}\nord =(z^{\frac1\lambdam}-w^{\frac1\lambdam})^{\alpha\cdot\beta}
\nord\eE^{\ii(\alpha\cdot\hat Q(z)+\beta\cdot\hat Q(w))}\nord.\ee
We emphasize in particular that the integral affine subalgebra
generated by $\Jahhr a 0=(\hat E_\alpha^{(0)},\Jahhr A 0)$, which
is an ordinary untwisted simply-laced affine algebra, is represented by this
construction at level $\hat k=\lambdam$.\\[2em]
\underline{\bf Cosets, conformal embeddings at level $\lambdam$ and so on}
\\[2em]
The $\g/\h$ coset construction \cite{baha,halp,goko,goko2} has the stress 
tensor $T_{\g/\h} = T_{\g} - T_{\h}$ and Virasoro central charge
$c_{\g/\h} = c_{\g} - c_{\h}$, where $\h$ is a simple subalgebra of
the simple\footnote{ The orbifold coset constructions are easily extended to
reductive $\g=\oplus \g_I$ and $\h=\oplus \h_i$ with levels $k_I(\g_I)$
and $k_i(\h_i)$, which results in orbifold currents at levels
$\hat k_I=\lambdam k_I$ and $\hat k_i=\lambdam k_i$ respectively.
}
 \lie\ $\g$ at level $k$. The image of this coset construction in 
the twisted sector of the corresponding cyclic orbifold is the set of 
orbifold stress tensors
\be \begin{array}{l}
\hat T^{(r)}_{\g/\h} = \hat T^{(r)}_{\g} - \hat T^{(r)}_{\h}
,~\,\,\,r=0\ldots\lambdam-1  \\[1em]
\ch_{\g/\h} = \ch_{\g} - \ch_{\h}=\lambdam c_{\g/\h} \, ,
\end{array}\ee
which satisfy the local form \erf{3.27} of the orbifold Virasoro algebra
with central charge $\ch_{\g/\h}$.
Here, ${\hat T}^{(r)}_{\g}$ 
(at level $\hat k=\lambdam k$ of the orbifold affine algebra)
is given in 
\erf{3.46},
and ${\hat T}^{(r)}_{\h}$ has the same form with $L^{ab}_\g\to L^{ab}_\h$.

The conformal embeddings \cite{baha,halp3,gono,scwa,babo} are the unitary coset
constructions with $c_{\g/\h} =0$ and hence $T_{\g} = T_{\h}$. 
In the mother theory, these
embeddings are known to occur only at level one of the ambient affine \lie.
It follows that each of these conformal embeddings has its image in
the corresponding cyclic orbifold
\be
\hat T_\g^{(r)}=\hat T_\h^{(r)},\,\,\,\,\,r=0\ldots\lambdam-1
\ee
at $\ch_{\g/\h}=0$ and level $\hat k=\lambdam$ of the ambient orbifold 
affine algebra (and its integral affine subalgebra). 
These higher level conformal embeddings include the chiral stress
tensor of the twisted sector when $r=0$.
We mention in particular 
the orbifold image of a familiar conformal embedding, whose form
\be \Thhr r (\mbox{simply-laced}\,\g_\lambdam) = \Thhr r
(\mbox{Cartan(simply-laced}\,\g\mbox{)}),\,\,\,\,\,\,r=0\ldots\lambdam-1 \ee
can also be verified directly from the higher-level orbifold vertex operator
construction (3.61-62). 

Beyond the coset constructions, one has the general affine-Virasoro 
construction \cite{haki,mprst,Hkoc}
\be
T(z)=L^{ab}\nord J_a(z)J_b(z)\nord,\,\,\,\,\,\,c=2k\eta_{ab}L^{ab}\ee
where $L^{ab}$ is any solution of the Virasoro master equation.
The image of the general affine-Virasoro construction in the twisted sector
of the corresponding cyclic orbifold is the set of  
orbifold stress tensors
\be
\Thhr r(z)=\frac1\lambdam L^{ab}\sum_{s=0}^{\lambdam-1}
\nord \Jahhr a s (z)\Jahhr b {r-s}(z)\nord\,+\,
\frac{\hat c}{24\lambdam z^2}(1-\frac1{\lambdam^2}) \,\delta_{r,0}
\labl{3.67new}
which satisfy the local form \erf{3.27} of the orbifold Virasoro algebra
with $\hat c=\lambdam c$. The normal ordering 
of the orbifold currents is given in eq. \erf{3.36}.

\subsection{Ising orbifolds at $\hat c=\frac \lambdam 2$}
As another example of the induction procedure, we construct the twisted 
sectors of Ising orbifolds at $\hat c=\frac\lambdam2$, starting from
the mother theory at $c=\frac12$. This example has the induction of
a Ramond sector as an additional feature.\\[2em] 
\underline{\bf 
$\widehat{BH/NS}$ sector}\\[2em]
The induction of this sector of the Ising orbifolds begins with the
$BH/NS$ sector \cite{baha,nesc} of the critical Ising model
\be T_{BH/NS}=-\frac12\nord H\partial H\nord\,,\,\,\,\,\,c=\frac12\ee
\be H(z)=H(z\eE^{2\pi\ii})=\sum_{m \in \zet} 
b(m+\frac12)z^{-(m+\frac12)-\frac12}\,.\ee
The orbifold induction procedure maps the $BH/NS$ sector of the mother
theory into the $\widehat{BH/NS}$ sector of the Ising orbifold,
where the principal primary fields corresponding to the 
$BH/NS$ fermion have the form
\be \bhr  r{m+\frac {r+\frac12}\lambdam}=\sqrt\lambdam \,b(\lambdam 
m+r+\frac12) ,\,\,\,\,\,r=0\ldots\lambdam-1\ee
\be \Hhhr r (z)=\rho^{\frac12}(z)\sum_{s=0}^{\lambdam-1}H(z^{\frac1\lambdam}
\eE^{\frac{2\pi\ii s}\lambdam})\eE^{\frac{2\pi\ii s}\lambdam(r+1)}=
\sum_m \bhr  r{m+\frac {r+\frac12}\lambdam}z^{-(m+\frac {r+\frac12}\lambdam)
-\frac12}\ee
\be
\Hhhr r (z\eE^{2\pi\ii})=\Hhhr r (z)
\eE^{-2\pi\ii(\frac{r+\frac12}\lambdam+\frac12)} \ee
\be 
\Hhhr r(z)\Hhhr s(w)=\frac{\lambda\delta_{r+s+1,\,0\bmod\lambdam}}{z-w}+\reg.
\labl{3.53}
\be
\bra 0 \Hhhr r (z) \Hhhr s (w) \mket 0
=(\lambdam \delta_{r+s+1,\,0\bmod\lambdam}) (\frac wz)^{\frac12
(\frac1\lambdam-1)+\frac r\lambdam}\frac1{z-w}\,.
\labl{<hh>}
The orbifold two-point correlators in \erf{<hh>} are examples of the
more general result \erf{<ff>}.
The generators of the orbifold Virasoro algebra in the 
$\widehat {BH/NS}$ sector are
\be\begin{array}{lll}
\Lvr r {m+\frac r\lambdam} &=& \frac 1{2\lambdam}
\sum_{s=0}^{\lambdam-1}\sum_n(m-n+\frac{r-s-\frac12}\lambdam) 
\nord \bhr s{n+\frac {s+\frac12}\lambdam} 
\bhr {r-s-1}{m-n+\frac{r-s-1+\frac12}\lambdam}\nord\\[1em]&&+
\delta_{r,\,0}\frac1{48}(\lambdam-\frac1\lambdam) 
,\,\,\,\,\,\,\,\,\,\,\,r=0\ldots\lambdam-1
\end{array}\ee
\be\begin{array}{lll}
\nord \bhr r {m+\frac{r+\frac12}\lambdam}
\bhr s {n+\frac{s+\frac12}\lambdam} \nord&\equiv&
-\theta(m+\frac{r+\frac12}\lambdam>0)\,\bhr s {n+\frac{s+\frac12}\lambdam}
\bhr r {m+\frac{r+\frac12}\lambdam}\\[1em]&&
+\theta(m+\frac{r+\frac12}\lambdam<0)\,\bhr r {m+\frac{r+\frac12}\lambdam}
\bhr s {n+\frac{s+\frac12}\lambdam}\,\end{array}\ee
with central charge $\hat c =\frac \lambdam2$.
The $BH/NS$ vacuum $\mket 0$ of the mother theory is now the ground state of 
the 
twisted sector, which is the twist field of the orbifold. It is a primary 
field under the 
integral Virasoro subalgebra $\Lvr 0 m$ with conformal weight 
\be\hat\Delta=\frac1{48}(\lambdam-\frac1\lambdam)\,.\ee 
The principal primary states created by the orbifold fermion fields 
$\Hhhr r(z)$ 
are primary with conformal weights 
\be
\hat\Delta_r=1+\frac\lambdam{48}-\frac1\lambdam(r+\frac{25}{48})
,\,\,\,\,\,r=0\ldots\lambdam-1
\ee
under the integral Virasoro subalgebra of the orbifold.

\pagebreak

\underline{\bf $\widehat R$ sector}\\[2em]
To induce the $\widehat R$ sector of the Ising orbifolds, we begin 
with the $R$ sector \cite{ramo} of the critical Ising model
\be T_R=-\frac12\nord H\partial H\nord+\frac1{16z^2},\,\,\,\,\,c=\frac12\ee
\be H(z)=\sum_{m\in\zet} b(m)z^{-m-\frac12}\ee
\be H(z\eE^{2\pi\ii})=-H(z),\,\,\,\,\,\theta=\frac12 \labl{3.60}
In this case, the monodromy $\theta$  of the $R$ fermion in \erf{3.60} is a 
constant. So long as we do not assume that 
$\theta$ annihilates the ground state, the operator formalism of Subsection 
3.1 is still applicable.
We obtain the orbifold fermion fields of the $\widehat R$ sector 
\be \bhr  r{m+\frac r\lambdam}=\sqrt\lambdam \, b(\lambdam m+r)
,\,\,r=0\ldots\lambdam-1\ee
\be \Hhhr r (z)=\rho^{\frac12}(z)\sum_{s=0}^{\lambdam-1}H(z^{\frac1\lambdam}
\eE^{\frac{2\pi\ii s}\lambdam})\eE^{\frac{2\pi\ii s}\lambdam(r+\frac12)}=
\sum_m \bhr  r{m+\frac r\lambdam}z^{-(m+\frac r\lambdam)
-\frac12}\ee
\be
\Hhhr r (z\eE^{2\pi\ii})=
\Hhhr r (z)\eE^{-2\pi\ii(\frac r\lambdam+\frac12)} \ee
\be 
\Hhhr r(z)\Hhhr s(w)=\frac{\lambda\delta_{r+s,\,0\bmod\lambdam}}{z-w}+\reg.
\labl{3.64}
The generators of the orbifold Virasoro algebra 
in the $\widehat R$ sector are
\be\begin{array}{lll}
\Lvr r {m+\frac r\lambdam} &=& \frac 1{2\lambdam}
\sum_{s=0}^{\lambdam-1}\sum_n(m-n+\frac{r-s}\lambdam) 
\nord \bhr s{n+\frac s\lambdam} 
\bhr {r-s}{m-n+\frac{r-s}\lambdam}\nord\\[1em]&&+\delta_{m+\frac 
r\lambdam,\,0} (\frac1{16\lambdam}+\frac1{48}(\lambdam-\frac1\lambdam)) 
,\,\,\,\,\,\,r=0\ldots\lambdam-1
\end{array}\ee 
\be\begin{array}{lll}
\nord \bhr r {m+\frac r\lambdam}
\bhr s {n+\frac s\lambdam} \nord&\equiv&
-\theta(m+\frac r\lambdam>0)\,\bhr s {n+\frac s\lambdam}
\bhr r {m+\frac r\lambdam}\\[1em]&&
+\frac12\delta_{m+\frac r\lambdam,\,0}\,[\bhr 00,\bhr s {n+\frac s\lambdam}]
\\[1em]&& +\theta(m+\frac r\lambdam<0)\,\bhr r {m+\frac r\lambdam}
\bhr s {n+\frac s\lambdam}\,\end{array}\ee
with central charge $\hat c =\frac \lambdam2$.
The ground state $\mket r$ of the mother Ramond sector becomes the ground 
state of the $\widehat R$ sector of the orbifold.  It is a primary state 
with conformal weight 
\be \hat\Delta=\frac1{24}(\frac\lambdam 2+\frac1\lambdam)\,\ee
as measured by the integral Virasoro subalgebra $\Lvr 0 m$.

\subsection{Orbifold SL(2,$\reals$) Ward identities}
As discussed in Subsection 2.7, the orbifold Virasoro algebra has a 
centrally-extended $\SltR$ subalgebra (see \ref{sltr}) which is generated by 
\be \Lvr 1 {\frac1\lambdam},\,\,\Lvr 0 0,\,\,
\Lvr {\lambdam-1}{-1+\frac{\lambdam-1}\lambdam} \,.\labl{3.85}
These generators do not form a subalgebra of the 
integral Virasoro 
subalgebra of the orbifold, and moreover, eq. \erf{2.32} shows that 
the ground state $\mket 0$ of the twisted sector is not
annihilated by these generators. Nevertheless, we 
find the {\it orbifold} $\,\SltR$ {\it Ward identities}
\be \begin{array}{c}
\bra 0 [\Lvr 1 {\frac 1 \lambdam},A]\mket 0 =\bra 0 [\Lvr 0 0,A]
\mket 0 = 0 \\[2em]
\bra 0 [\Lvr {\lambdam-1} {-1+\frac {\lambdam -1} \lambdam},A]\mket 0
= 0
\end{array}\labl{3.69}
in the twisted sector, where $A$ is any operator. These Ward identities
are the images in the orbifold of the $\SltR$ Ward identities 
\cite{bepz} of the mother theory.

As an application, we will work out the explicit form of the orbifold 
Ward identities \erf{3.69} in the case of the principal primary fields.
For this, we need the commutators
\be \begin{array}{rll}
[\Lvr 1 {\frac 1 \lambdam},\phidhhr s (z)]&=&z^{\frac 1 \lambdam}
(z\delz +\Delta(1+\frac 1 \lambdam ))\phidhhr {s+1} (z) \\[2em]
[\Lvr 0 0,\phidhhr s (z)]&=&(z\delz +\Delta)\phidhhr {s} (z) \\[2em]
[\Lvr {\lambdam-1} {-1+\frac {\lambdam-1} \lambdam},\phidhhr s 
(z)]&=&z^{-\frac 1\lambdam}(z\delz +\Delta(1-\frac 1 \lambdam 
))\phidhhr {s-1} (z) 
\end{array}\labl{q2}
which follow from the more general relation \erf{2}.
Then we find the orbifold $\SltR$ Ward identities 
\be A_{\Delta_1...\Delta_n}^{s_1...s_n}(z)\equiv
\bra 0 \phidnhhr 1 
{s_1}(z_1)... \phidnhhr n {s_n} (z_n) \mket 0 \labl{3.72}
\be \begin{array}{lll}
0&=&\sum_{i=1}^n z_i^{\frac 1 \lambdam} (z_i\partial_i+\Delta_i(1+\frac 
1 \lambdam))A_{\Delta_1...\Delta_i...\Delta_n}^{s_1...s_i+1...s_n}(z)\\[2em]
0&=&\sum_{i=1}^n (z_i\partial_i+\Delta_i) 
A_{\Delta_1...\Delta_i...\Delta_n}^{s_1...s_i...s_n}(z) \\[2em]
0&=&\sum_{i=1}^n z_i^{-\frac 1 \lambdam} (z_i\partial_i+\Delta_i(1-\frac 
1 \lambdam))A_{\Delta_1...\Delta_i...\Delta_n}^{s_1...s_i-1...s_n}(z) \, 
\end{array}\labl{3.73}
for the matrix correlators defined in \erf{3.72}. Since $\hat 
\varphi^{(s)}$
is cyclic in $s$, these correlators are cyclic in each index $s_i$.

The matrix differential equations \erf {3.73} are difficult to solve in 
general, although there are some simple observations about the solutions 
which are easily made:\\ 
$\bullet$~~ Because $\Lvr 0 0$ is the scale operator, the system 
\erf{3.73} 
implies that the singularities of correlators and OPE's
in the twisted sectors of the orbifold are controlled by the conformal
weights of the fields in the mother theory. We have 
already seen this phenomenon 
in many examples (c.f. equations \ref{3.20}, \ref{<ff>}, \ref{3.26},
\ref{3.27}, \ref{3.40}, \ref{<jj>}, \ref{3.41}, \ref{3.53}, 
\ref{<hh>} and \ref{3.64}).\\
$\bullet$~~ Orbifold correlators are not translation invariant in the 
twisted sector (c.f. \ref{<ff>}):
In the first place, the operator $\Lvr {\lambdam-1}{-1+\frac{\lambdam-1}
\lambdam}\,$, which is the image of the translation operator in the mother
theory, is no longer the generator of translations (c.f. \ref{q2}). There
is, on the other hand, a translation operator $\Lvr 0 {-1}$ in the integral
Virasoro subalgebra, but this operator fails to annihilate the ground 
state of the twisted sector.\\
$\bullet$~~ The solution for the one-point correlators 
\be \bra 0 \phidhhr r (z) \mket 0 = 0,\,\,\,\,\,
r=0\ldots\lambdam-1
\labl{3.74}
follows easily from the system \erf{3.73}, and the result \erf{3.74} also 
follows directly 
from eq. \erf{de1} because $\langle\phid\rangle=0$ in the mother theory.
The orbifold two-point correlators in \erf{<ff>}, \erf{<jj>} and
\erf{<hh>} are also  solutions of the system \erf{3.73}, but
these are only particular solutions corresponding
to trivial monodromy in the mother theory. Beyond this, the solutions of 
the system \erf{3.73} must be
quite complex: One expects the analogues of the usual 
ambiguities for $n\geq 4$, and, moreover,
the general solutions for $n\geq 2$ must also 
allow for the structure induced in the orbifold by the monodromy 
algebra of the mother theory (c.f. eq. \ref{3.22}).

\subsection{Orbifold null-state differential equations}
Under the orbifold induction procedure, any null state of the mother theory 
maps to a null state in the twisted
sector of the cyclic orbifold, and, as a consequence, every BPZ null-state
differential equation has its image in the corresponding orbifold.

As an example, we will focus on the cyclic orbifolds formed from tensor
products of the Virasoro minimal models \cite{bepz},
and in particular on the images of the simplest null states of 
the mother theories which now read in the orbifold
\be 
0 = \left( \lambdam (\Lvr{\lambdam-1}{-1+\frac{\lambdam-1}\lambdam})^2
- \frac23(1+2\Delta)\Lvr{\lambdam-2}{-1+\frac{\lambdam-2}\lambdam} \right)
\mket{\Deltah_{\lambdam-1}} 
\ee
\be 
\Deltah_{\lambdam-1}=\frac\Delta\lambdam+\frac{\hat c}{24}\,
(1-\frac1{\lambdam^2}),~~~~~~~\,\,\,\,\,\Delta = \left\{ \begin{array}{l}
\frac m{4(m+3)} \\[.5em] \mbox{or} \\[.5em]
\frac {m+5}{4(m+2)} \end{array}\right.  
\ee
\be \ch = \lambdam (1-\frac6{(m+2)(m+3)} ),\,\,\,m\geq 0 \, . \ee
Following the usual procedure, one then obtains the following null-state
matrix differential equations
\be B_\Delta^{r_1 \ldots r_{n-1}} (z) \equiv
\bra{0} \phidohhr{r_1}(z_1) \ldots \phidnmohhr{r_{n-1}}(z_{n-1})
\mket {\Deltah_{\lambdam-1}} \labl{3.94}
\be\begin{array}{lll}
0&=& \sum_{i=1}^{n-1} \{ \lambda
(z_i^{1-\frac1\lambdam}\partial_i + \frac{\Delta_i(1-\frac1\lambdam)}
{z_i^{\frac1\lambdam}} )^2 
+ \frac23 (1+2\Delta) 
(z_i^{1-\frac2\lambdam}\partial_i + \frac{\Delta_i(2-\frac1\lambdam)}
{z_i^{\frac2\lambdam}} ) \}
B_\Delta^{r_1 \ldots r_i-2 \ldots r_{n-1}}(z)  \\[2em]
&& +  2\lambdam \sum_{i<j}
(z_i^{1-\frac1\lambdam}\partial_i + \frac{\Delta_i(1-\frac1\lambdam)}
{z_i^{\frac1\lambdam}} )
(z_j^{1-\frac1\lambdam}\partial_j + \frac{\Delta_j(1-\frac1\lambdam)}
{z_j^{\frac1\lambdam}} )
B_\Delta^{r_1 \ldots r_i-1 \ldots r_j-1 \ldots r_{n-1}}(z) 
\end{array}\ee
for the correlators defined in \erf{3.94}.
Here $\hat\varphi_{\Delta_i}^{(r_i)}$  are the principal primary fields 
corresponding to the Virasoro primary fields $\varphi_{\Delta_i}$ of
conformal weight $\Delta_i$ in the minimal model.
Because there is no translation invariance in the twisted 
sector of the orbifold, we have 
been unable to express this system in a more familiar form.

\subsection{The conventional view of orbifolds}

In this subsection we want to explain in further detail why the twisted
sectors discussed above are in fact the twisted sectors of cyclic orbifolds.
The discussion will also shed further light on the 
higher-twist components of the orbifold algebras of Section 2.
To this end, we first review the more conventional viewpoint of orbifolds.
\\[2em]
\underline{\bf General orbifolds}
\\[2em]
At this point it is helpful to recall a few standard facts about the
algebraic description of orbifolds (see e.g.\ Ref.\ \cite{dvvv}): Given a
\cft\ with a chiral algebra $\cala$ and a symmetry of $\cala$, the idea
of the orbifold construction
is to consider the subalgebra $\calao$ of $\cala$ that is fixed under
the symmetry. The subalgebra $\calao$ is the chiral algebra of a new \cft, 
called the orbifold theory. It is a well-known result \cite{mose3} in \cft\ 
that, in order to obtain a modular invariant partition function,
we should find {\em all} irreducible \rep s of $\cala^{(0)}$ and retain  
each inequivalent \rep\ once.

Some \rep s of $\calao$ are easy to obtain: Any \irrep\ of the original
chiral algebra $\cala$ can be decomposed into \irrep s of $\calao$. 
The states in these $\calao$-modules form what is called the untwisted
sector of the orbifold theory. An important feature of the untwisted sector 
is that
the \rep\ theory of the original chiral algebra $\cala$ provides us with a good
handle on this sector. In the process of this decomposition, two
effects can occur, and we will encounter both of them: \\
$\bullet$~~ Irreducible modules of $\cala$ can be reducible under $\calao$. \\
$\bullet$~~ Irreducible modules which are inequivalent as modules of $\cala$
            can be isomorphic as modules of the subalgebra $\calao$.

Generically, however, there are modules of $\calao$ which are not contained
in any $\cala$ module. The states in these modules form what is called the
twisted sector of the orbifold. This sector usually causes problems in the
description of orbifolds, since the \rep\ of $\cala$ does not provide a handle
on it. However, as we will emphasize shortly, in the case of cyclic permutation
orbifolds, the orbifold induction procedure of Sections 2 and 3 has given 
us in fact an explicit construction of the states in the twisted sector.

To further understand the twisted sector, we briefly review a few more facts 
about orbifolds in their geometric description. For simplicity, we will 
assume that the symmetry group $G$ of $\cala$ by which we mod out is abelian. 
Let us further imagine that the theory in question can be constructed from
string coordinates $X(\tau,\sigma)$, on which $G$ acts non-trivially. The 
worldsheet is parametrized by $\tau$ and $\sigma$, where 
$\sigma+2\pi \equiv \sigma$. If one mods out by $G$,
one must also introduce twisted sectors \cite{dhvw,dhvw2}, 
in which the string closes only up to the action of $G$,
\be X(\tau,\sigma+2\pi) = g X(\tau,\sigma) \,  \labl{orb}
where $g$ is an element of $G$. This relation tells us that the string
coordinate in the twisted sectors is a multi-valued function on the world-sheet.
This is the origin of root coverings of the worldsheet in the 
calculation of orbifold $N$-point functions in examples \cite{hava}  and it is
also the origin of the root-covering algebras in Section 2.

To translate this prescription into a more algebraic form, we introduce the
complex variable $z=\exp(\tau+\ii \sigma)$ which maps the worldsheet
cylinder to the complex plane, where time flows radially.
In these coordinates, the requirement \erf{orb} translates into a 
monodromy property
\be X(z \mono) = g X(z) \labl{720}
of the string coordinate. After quantization, the string coordinate becomes
an operator and the modes of $\partial X$ contain both twisted and untwisted
Cartan currents; The untwisted Cartan currents form the \rep\ of the
orbifold chiral algebra in the twisted sector, while the twisted modes are the
twisted sector representatives of the other generators of the original
chiral algebra $\cala$, which now play the role of intertwiners in the
twisted sector.

Many of the principles discussed in this subsection will be seen again in
our discussion of orbifolds on the torus (cf.\ Section 4).
\\[2em]
\underline{\bf Cyclic orbifolds}
\\[2em]
We now specialize this discussion to the case of cyclic permutation
orbifolds. These orbifolds are constructed by modding out the symmetry
$\zet_{\lambdam}$ which acts by cyclic permutations
on the tensor product of $\lambdam$ copies of the
mother theory, so that they have Virasoro central charge $\ch=\lambdam c$,
where $c$ is the central charge of the mother theory.

As is true in general orbifolds, the orbifold chiral algebra $\calao$ of the
cyclic orbifold is a subalgebra of the original chiral algebra of the 
tensor product theory.
The subalgebra $\calao$ is represented in the
twisted sector by the $r=0$ component of the orbifold affine and Virasoro
algebras of Section 2. The rest of the original chiral
algebra still operates (as an abstract algebra) in the twisted sectors of 
the orbifold, but it is represented in the twisted sectors by operators with 
\nontriv\ monodromy. These operators are the higher-twist components
$r=1\ldots\lambdam-1$ of the orbifold affine and Virasoro algebras of
Section 2. 

Let us illustrate this in the simplest case, when $\lambdam=2$ and the \cft\
is a tensor product of two affine-Sugawara constructions at level $k$.
The symmetry by which we want to mod out is the permutation
of the two factors in the tensor product. In the untwisted sector, we have a 
set of operators in $\cala$ that are even under the $\zet_2$ permutation
\be J_a(z)\otimes\bfe + \bfe\otimes J_a(z) \, \labl{3106}
which span an affine \lie\ at level $2k$. Since the symmetry operates 
trivially on operators of the form \erf{3106}, then, according to \erf{720}, the abstract
algebra generated by these operators should be represented in the twisted 
sector of the orbifold by integer moded operators, 
\be J_a(z)\otimes\bfe + \bfe \otimes J_a(z) \quad \to \quad
\Jahr a 0 z \,  \ee
where $\Jahr a 0 z$ generate the integral affine subalgebra of the $\lambdam=2$
orbifold affine algebra \erf{2.1}. The odd part of $\cala$ in the
untwisted sector is
\be J_a(z)\otimes\bfe - \bfe\otimes J_a(z) \, \ee
and it transforms in the adjoint \rep\ with respect to the even part $\calao$.
According to \erf{720}, the odd part of $\cala$ must be represented in the
twisted sector by half-integer moded operators
\be J_a(z)\otimes\bfe - \bfe \otimes J_a(z) \quad \to \quad
\Jahr a 1 z \,  \ee
where $\Jahr a 1 z$ is the set of $r=1$ generators of the $\lambdam=2$
orbifold affine algebra \erf{2.1}. 

In summary, the orbifold algebras are simply the
representations in the twisted sectors of the chiral algebra of the
mother theory; moreover, the $r=0$ subalgebra of the orbifold algebra is the
representative in the twisted sector of the chiral algebra of the orbifold,
while the higher-twist components of the orbifold algebra have become
intertwiners in the twisted sector.

\sect{Characters and modular transformations}

So far our discussion has focussed on the orbifold \cft\ for a worldsheet of 
genus zero. We will now discuss some aspects of genus one, that is, 
the characters and their modular transformation properties. In what follows
we will, for the sake of simplicity, first discuss the case of $\lambdam=2$.

\subsection{The untwisted sector for $\lambdam=2$}

We begin with the untwisted sector. According to the discussion in the 
previous section, the states of the untwisted sector 
of the cyclic orbifold are obtained from states 
in the tensor product theory, which consists of $\lambdam$ copies of the 
mother \cft\ $\calc$. We assume here that
the mother theory is a rational \cft\ with Virasoro central charge $c \, $.
To each primary state with conformal weight $\Delta_i$ in this theory 
we associate the irreducible module $\hil_i$ spanned by the primary field
and its descendants with respect to the chiral algebra of $\calc$. We will
use characters to describe these modules:
\be \chi_i (\tau, z) \equiv  \Tr^{}_{\hil_i} \eE^{2\pi\ii\tau (\Lv 0-c/24) }
\eE^{2\pi\ii z \cdot J(0)} \, . \ee
In case the theory contains commuting spin one currents $J_a(z)$, we have also
included in the definition of the character a set of Cartan angles, collected
in the vector $z$.
The primary fields of the tensor product theory are labeled by $\lambdam$
indices, each of which labels a primary field of the mother theory.
The corresponding module is the tensor product of the modules of the mother 
theory and, correspondingly, the characters of the tensor product theory 
\be \chi_{ij}(\tau, z) = \chi_i(\tau,z)\chi_j(\tau,z) \,  \ee
are just the products of the characters of the mother theory.
\\[2em]
\underline{\bf Off-diagonal fields}\\[2em]
Let us now describe the characters of the primary fields in the untwisted 
sector of the orbifold theory. If $i$ and $j$ are different, the direct sum
$(\hil_i \otimes\hil_j)\, \oplus\, (\hil_j\otimes\hil_i)$ carries a \rep\
of the $\zet_2$ permutation symmetry, which we use to decompose the direct
sum into two submodules of the orbifold chiral algebra: One of these
submodules contains the states which are symmetric under the permutation
and the other contains the antisymmetric states. 
The symmetric and antisymmetric states are of the form
\be
\mket x\otimes \mket y + \mket y \otimes \mket x,
~~~~
\mket x\otimes \mket y - \mket y \otimes \mket x
\ee
respectively,
where $\mket{x} \in \hil_i$ and $\mket y\in\hil_j$. However,
the two modules are isomorphic: Symmetric and antisymmetric
states come always in pairs, and the mapping
\be
\mket x\otimes \mket y + \mket y \otimes \mket x ~\to~
\mket x\otimes \mket y - \mket y \otimes \mket x
\ee
is an
intertwiner of the \rep s. In the orbifold theory we should mod out by this
permutation and we should keep only one of these two modules.
Hence we retain only a single primary field called $(ij)$ in the untwisted 
sector of the orbifold theory with character 
\be \Chi_{(ij)}(\tau, z) = \chi_i(\tau,z)\chi_j(\tau,z) \, , \, 
\quad \,\, \, i< j \, . \labl{43}
Characters of the orbifold theory will be denoted by the symbol 
$\Chi$.\\[2em]
\underline{\bf Diagonal fields}\\[2em]
The diagonal fields for which $i=j$ also split into two 
\rep s, symmetric and antisymmetric under the orbifold chiral algebra. 
We will present strong evidence\footnote{ Using the techniques developed 
in \cite{dolm5} it should be 
possible to prove directly that these modules are irreducible.} 
that these \rep s are indeed irreducible \rep s of the orbifold chiral
algebra: We will compute the characters of the corresponding modules and
show that they span a module of the modular group $SL(2,\zet)$. 
We denote the diagonal 
primary fields by $(i,\psi)$, where $\psi$ takes the values $0$ for
the symmetric and $1$ for the antisymmetric \rep. 

To compute the characters of these two primary fields, we remark that 
the permutation $\pi$ gives rise to an involution $T_\pi$ on 
$\hil_i \otimes \hil_i$ which maps
the state $\mket x \otimes \mket y$ to the state $\mket y \otimes \mket x$. 
We introduce the character-valued index
\be \Chi^\pi_i(\tau,z) \equiv \Tr^{}_{\hil_i\otimes \hil_i} T_\pi 
\eE^{2\pi\ii\tau(\Lv 0-\ch/24)}\eE^{2\pi\ii z\cdot J(0)} \,  \labl{45}
where $\Lv 0$ is the zero mode of the Virasoro algebra of the tensor
product theory and $\ch=\lambdam c$ is the central charge of the tensor
product theory. The index $\Chi^\pi_i$
is a close relative of the twining characters introduced in \cite{fusS3}:
It encodes the action of an outer automorphism of the chiral algebra
on the space of states, in this case the action of the permutation symmetry.
In the case at hand the index is easy to compute: 
Only states of the form $\mket x\otimes\mket x$ 
contribute to $\Chi^\pi_i$, and evaluation of their contribution gives
\be \Chi^\pi_i(\tau,z) = \chi_i(2\tau,2z)\, \ee
since the eigenvalues of $\Lv 0$ and $J(0)$ are additive.

Because $\frac12(\bfe\pm T_\pi)$ is the projection operator on 
symmetric and antisymmetric states respectively in $\hil_i\otimes \hil_i$, we
may then evaluate the characters of the symmetric and antisymmetric parts
separately:
\be\begin{array}{lll}
 \Chi_{(i,\psi)}(\tau,z) &=& \Tr^{}_{\hil_i\otimes \hil_i} \frac12 (\bfe 
+ \eE^{2\pi\ii \psi/2} T_\pi)
\eE^{2\pi\ii\tau(\Lv 0-\ch/24)}\eE^{2\pi\ii z \cdot J(0)} 
= \frac12\chi_i(\tau,z)^2 + \eE^{\pi\ii \psi} \frac12\Chi^\pi_i(\tau,z) \\[1em]
&=& 
\frac12\chi_i(\tau,z)^2 + \eE^{\pi\ii \psi} \frac12\chi_i(2\tau,2z) \, . 
\end{array}\labl{46}
It is easy to see that this field content reproduces the partition function
of the untwisted sector derived in \cite{klsC}:
\be \begin{array}{lll}
Z_{{\rm untwisted}} (\tau, z) 
&\equiv & \sum_{i<j} | \Chi_{(ij)}(\tau,z)|^2 + \sum_i
|\Chi_{(i,\,0)}(\tau,z)|^2 +|\Chi_{(i,1)}(\tau,z)|^2 \\[1em]
&=& \frac12 \sum_{i,j} |\chi_i(\tau,z)|^2 |\chi_j(\tau,z)|^2
+ \frac12 \sum_i |\chi_i(2\tau,2z)|^2 \,  \end{array}\labl{pfn}
where we have used equations \erf{43} and \erf{46} to obtain the final
form in \erf{pfn}.

For later use we also read off the modular matrix $T$ describing the
modular transformation $\tau\to\tau+1$ in the untwisted sector: It is
a diagonal matrix with entries
\be T_{(ij)} = T_i T_j\, , \quad T_{(i,\psi)} = T_i^2 \, . \ee
Here $T_i = \exp(2\pi\ii (\Delta_i-\frac c{24}))$ are the diagonal elements
of the modular matrix $T_{ij} = \delta_{ij} T_i$ of
the mother conformal field theory.

\subsection{The untwisted sector for general $\lambdam$}

The untwisted sector of cyclic permutation orbifolds for general $\lambdam$
can be described quite similarly, except that the notation becomes somewhat more
complicated. We describe fields by multiindices $\vec i=(i_1,i_2,\ldots, 
i_\lambdam)$, which generalizes the notation $(ij)$ in the previous 
subsection.
Correspondingly, we introduce the following notation for the modules of
the tensor product theory:
\be \hil_{\vec i} \equiv \hil_{i_1}\otimes\hil_{i_2}\otimes\ldots\otimes
\hil_{i_\lambdam}\, . \ee
The action of the group of cyclic permutations of indices organizes
the multiindices into orbits; to each orbit $\vec i$ we associate its
stabilizer $\calS_{\vec i}$, the subgroup of elements of $\zet_\lambdam$ that 
leaves $\vec i$ fixed. Note that the stabilizer subgroup
does not depend on the choice of representative of the orbit, since 
$\zet_\lambdam$ is abelian.

Suppose now that $\pi\in \calS_{\vec i}$. Define $b(\pi)$ to be the order of 
$\pi$ and $a(\pi)=\lambdam/b(\pi)$.
As we saw for the diagonal fields in the case $\lambdam=2$, we can implement 
the action of $\pi$
on the corresponding module and introduce analogous character-valued
indices. By the same reasoning, we obtain
\be\begin{array}{lll}
\Chi^\pi_{\vec i}(\tau,z) 
&\equiv& \Tr^{}_{\hil_{\vec i}} 
T_\pi \eE^{2\pi\ii\tau(\Lv 0 -\ch/24)}\eE^{2\pi\ii z \cdot J(0)} \\[1em]
&=& \prod_{l=1}^{a(\pi)} \chi_{i_l}(b(\pi)\tau, b(\pi) z) 
\end{array}\ee
Again, we have to decompose the irreducible module of the tensor product theory 
into submodules of the orbifold chiral algebra. These submodules are obtained 
by decomposing $\hil_{\vec i}$ into subspaces on which all $T_\pi$ act 
diagonally, when $\pi$ is in the stabilizer. The decomposition 
\be \hil_{\vec i} = \oplus_\psi \hil_{(\vec i,\psi)} \ee
should be such that on each subspace $\hil_{(\vec i,\psi)}$
any $\rho\in\calS_{\vec i}$ acts as a phase 
$\eE^{2\pi\ii\psi m(\rho) /\lambdam}\bfe$. 
Here we identify $\rho\in\calS_{\vec i}$ with an integer
$m(\rho)\in \{0,\ldots ,\lambdam-1\}$, and $\psi$ is an integer ranging
from $0$ to $|\calS_{\vec i}|-1$, where $|\calS_{\vec i}|$ is the number 
of elements in the stabilizer $\calS_{\vec i}$.
In other words, the true characters are labeled by a multiindex $\vec i$ and 
an element $\psi$ of the character group of the stabilizer. 

To compute the characters of the spaces $\hil_{(\vec i,\psi)}$ separately,
we remark that 
\be \Chi^\rho_{\vec i}(\tau,z) = \sum_{\psi=0}^{|\calS_{\vec i}|-1} 
\Tr^{}_{\hil_{(\vec i,\psi)}} T_\rho 
\eE^{2\pi\ii\tau(\Lv 0-\ch/24)}\eE^{2\pi\ii z \cdot J(0)} 
= \sum_{\psi=0}^{|\calS_{\vec i}|-1} \eE^{2\pi\ii\psi m(\rho)/\lambdam} 
\Tr^{}_{\hil_{(\vec i,\psi)}}  
\eE^{2\pi\ii\tau(\Lv 0-\ch/24)}\eE^{2\pi\ii z \cdot J(0)} \, . \ee
Inverting this relation, we find 
\be \Chi_{(\vec i,\psi)}(\tau,z) = \Tr^{}_{\hil_{(\vec i,\psi)}}
\eE^{2\pi\ii\tau(\Lv 0-\ch/24)}\eE^{2\pi\ii z \cdot J(0)} =
\frac1{|\calS_{\vec i}|} \sum_{\rho\in S_{\vec i}} 
\eE^{-2\pi\ii \psi m(\rho)/\lambdam} \Chi^\rho_{\vec i}(\tau,z) 
\labl{810}
for the characters of the primary fields
in the untwisted sector.
\\[2em]
\underline{\bf Prime $\lambdam$}\\[2em]
Let us briefly discuss the special case when $\lambdam$ is prime. In this
case the stabilizer is either trivial or $\zet_\lambdam$. In the first case,
we have $\calS_{\vec i}=\{e\}$ and 
the character is simply a product of the characters of the mother theory:
\be \Chi_{(\vec i, 0)}(\tau,z) = \Chi_{\vec i}^e(\tau,z)
= \prod_{l=1}^\lambdam \chi_{i_l}(\tau,z) \, . \ee
In case the stabilizer is non-trivial (which corresponds to the diagonal fields
in the previous subsection), we find that the multiindex has to be of the 
form $\vec i =(i,i,\ldots,i)$. Moreover, we have
\be \Chi_{\vec i}^\pi(\tau,z)= \chi_i(\lambdam \tau,\lambdam z) \,  \ee
when $\pi\neq e$.

The general formula \erf{810} gives in the case $\psi=0$
\be \Chi_{(\vec i, 0)} (\tau,z) = \frac 1\lambdam (\chi_{i}(\tau,z) )^{\lambdam}
+ \frac{\lambdam-1}\lambdam \chi_{i}(\lambdam\tau,\lambdam z)\ee
while for $\psi\neq0$ we get 
\be \Chi_{(\vec i,\psi)}(\tau,z) 
= \frac 1\lambdam (\chi_{i}(\tau,z) )^{\lambdam}
- \frac1\lambdam \chi_{i}(\lambdam\tau,\lambdam z) \, . \labl{417}
Note that this result is independent of $\psi$ so that 
$\lambdam-1$ characters of the orbifold theory coincide.
As in equation \erf{pfn}, one can compute the partition function of the
untwisted sector and find full agreement with \cite{klsC}.

\subsection{The twisted sector for $\lambdam=2$}

Let us now turn to the twisted sectors. We claim that the characters of the
primary fields in the twisted sector can be obtained by considering the 
action of the orbifold chiral algebra on the modules of the original theory.
As a first step let us calculate (for arbitrary $\lambdam$):
\be \begin{array}{lll}
\chi_k(\frac{\tau+n}{\lambdam}, z) &=&
\Tr^{}_{\hil_k} \eE^{2\pi\ii\frac{\tau+n}\lambdam (\Lv0 -\frac c{24})} 
\eE^{2\pi\ii z \cdot J(0)}
= \Tr^{}_{\hil_k}\eE^{2\pi\ii\tau (\Lvr00 -\frac{\hat c}{24})} 
\eE^{2\pi\ii z\cdot\hat J(0)} \eE^{2\pi\ii n (\Lvr00 - \frac{\hat c}{24})} \, . 
\end{array}\labl{12}
Here $\Lvr00$ is the generator of the orbifold Virasoro algebra introduced
in \erf{2.5} and $\ch=\lambdam c$ is the Virasoro central charge of the
orbifold theory. The right side of equation \erf{12} shows that, up to
an extra $n$-dependent insertion, this is just the character 
of the module of the appropriate orbifold algebra.

The conformal weights of descendants of the same primary field differ only by
an integer, and we should therefore decompose the vector space $\hil_k$
into subspaces consisting of states that differ in conformal weights 
(with respect to $\Lvr00$) by integers. We denote these subspaces by 
$\hil_{\Widetilde{(k,\psi)}}$ and their characters by
$\Chi_{\Widetilde{(k,\psi)}}$. The fractional part  of the $\Lvr00$-conformal 
weight of states in $\hil_{\Widetilde{(k,\psi)}}$ is given by 
\be \Delta_{\Widetilde{(i,\psi)}} = \frac1\lambdam \Delta_i 
+\frac c{24}(\lambdam-\frac1\lambdam) + \frac\psi\lambdam  \labl{419}
with $\psi=0\ldots \lambdam-1$. Up to an integer contribution, this form 
is equivalent to the set
of conformal weights $\hat\Delta_r$ in \erf{states} of the principal primary
fields. As a consequence, the elements of the modular matrix $T$ for the
primary fields in the twisted sector read
\be T_{\Widetilde{(i,\psi)}} = \eE^{2\pi\ii(\Delta_{\Widetilde{(i,\psi)}}-
\ch/24)} \, , \ee
where $\Delta_{\Widetilde{(i,\psi)}}$ are the conformal weights in \erf{419}.
Equation \erf{12} shows us that 
\be \chi_k(\frac{\tau+n}{\lambdam}, z) = T_k^{n/\lambdam}
\sum_{\psi=0}^{\lambdam-1} \Chi_{\Widetilde{(k,\psi)}} (\tau,z) \, 
\eE^{2\pi\ii n\psi/\lambdam}\, , \labl{816}
where the quantity $T_k^{n/\lambdam}$ in \erf{816} is a particular root
of the modular matrix $T_k$ of the mother theory
\be T_k^{n/\lambdam}=
\exp(2\pi\ii\frac n\lambdam (\Delta_k -\frac c{24})) \, . \ee
Inverting relation \erf{816}, we find the characters of the primary fields
of the twisted sector
\be \Chi_{\Widetilde{(k,\psi)}}(\tau,z) 
= \Tr^{}_{\hil_{\Widetilde{(k,\psi)}}} \eE^{2\pi\ii \tau(\Lvr00 -\ch/24)}
\eE^{2\pi\ii z\cdot \hat J(0)} 
= \frac1\lambdam \sum_{n=0}^{\lambdam-1} \eE^{-2\pi\ii n\psi/\lambdam} 
T_k^{-n/\lambdam} \, \chi_k(\frac{\tau+n}{\lambdam}, z) \,  \labl{422}
where $k$ labels a primary field of the mother theory $\calc$ and 
$\psi=0\ldots\lambdam-1$. The relation \erf{422} is the analogue of the 
monodromy sums which were introduced earlier to decompose operators on the 
sphere, so the characters of the twisted sector are obtained by essentially
the same orbifold induction procedure presented in Sections 2 and 3. 

As a check on this result, we may compute a partition function for the
twisted sectors of the orbifold:
\be\begin{array}{lll}
Z'_{{\rm twisted}}(\tau,z) &=&
\sum_k \sum_{\psi=0}^{\lambdam-1} 
|\Chi_{\Widetilde{(k,\psi)}}(\tau,z)|^2 \\[1em]
&=& \sum_k
\frac1{\lambdam^2} \sum_\psi\sum_{n,n'=0}^{\lambdam-1} 
\chi_k(\frac{\tau+n}{\lambdam}, z)
\chi_k(\frac{\tau+n'}{\lambdam}, z)^*
T_k^{-n/\lambdam} T_k^{n'/\lambdam}
\eE^{-2\pi\ii n\psi/\lambdam} \eE^{2\pi\ii n'\psi/\lambdam} \\[1em]
&=&\sum_k \frac1\lambdam \sum_n |\chi_k(\frac{\tau+n}{\lambdam}, z)|^2 \, . 
\end{array}\labl{719}
The sum over $k$ in $Z'_{{\rm twisted}}$ 
is a sum over all primary fields in the 
mother theory $\calc$. Comparing with the known \cite{klsC} partition
function $Z_{{\rm twisted}}$ of the twisted sector of the orbifold,
we find that $Z_{{\rm twisted}} = (\lambdam-1) Z'_{{\rm twisted}}$. 
This shows that for $\lambdam=2$ the result $Z'_{{\rm twisted}}$ in \erf{719}
is the 
correct partition function, but for higher prime $\lambdam$ one must in fact
include $\lambdam-1$ copies of the same module of the orbifold algebra.
This parallels the situation found in the untwisted sector.
Before we explain this in more detail, we want to continue with
the case $\lambdam=2$, where this additional complication does not arise.

\subsection{Modular transformations for $\lambdam=2$}

Having computed the characters for the twisted and untwisted sectors at
$\lambdam=2$, we may now study the modular matrices $S^{{\rm orb}}$ and 
$T^{{\rm orb}}$ of the
orbifold. We have already determined the modular matrix $T^{{\rm orb}}$
which describes the behavior of the characters under $\tau\to\tau+1$. We now
compute the modular matrix $S^{{\rm orb}}$ which describes the 
transformation $\tau\to-\frac1\tau$. Via the Verlinde formula 
\be {\cal N}_{ijk}^{{\rm orb}} = \sum_n \frac{S^{{\rm orb}}_{in} 
S^{{\rm orb}}_{jn} S^{{\rm orb}}_{kn}}{S^{{\rm orb}}_{0n}} \labl{verl}
the modular matrices $S^{{\rm orb}}$
will also give the fusion rules of the orbifold theory. We denote by 
$0$ the vacuum $(k,\psi)=(0,0)$ of the orbifold theory; this is not the
ground state $\Widetilde{(k,\psi)} = \Widetilde{(0,0)}$ of the twisted sector
of the orbifold, which was called $\mket 0$ in Subsection 2.7.

To compute the $S$-matrix elements, we begin with the characters \erf{43}
of the off-diagonal fields in the untwisted sector. One finds that 
\be
\Chi_{(ij)}(-\frac1\tau,\frac z\tau) = 
\sum_{p<q} (S_{ip}S_{jq} + S_{iq}S_{jp}) \Chi_{(pq)}(\tau,z) 
+\sum_p S_{ip}S_{jp} \sum_{\psi=0}^1 \Chi_{(p,\psi)}(\tau,z)  , \ee
which implies the following orbifold $S$-matrix elements, 
\be \begin{array}{lll}
S_{(ij),(pq)} &=& S_{ip}S_{jq} + S_{iq}S_{jp} \\[.8em]
S_{(ij),(p,\psi)} &=& S_{ip}S_{jp} \\[.8em]
S_{(ij),\Widetilde{(p,\chi)}} &=& 0 \,\, . 
\end{array}\ee
Similarly, for the characters \erf{46} of the diagonal fields, we find that 
\be\begin{array}{lll}
\Chi_{(i,\psi)}(-\frac1\tau,\frac z\tau)&=&
\frac12 \sum_{p<q} S_{ip} S_{jq} \chi_p(\tau,z) \chi_q(\tau,z) 
+ \frac12 \eE^{2\pi\ii \psi/2} \sum_p S_{ip} \chi_p(\frac\tau 2,z) \\[1em]
&=& \sum_{p<q} S_{ip} S_{jq} \Chi_{(pq)}(\tau,z) 
+ \sum_p \frac12S_{ip}^2 \sum_{\psi=0}^1 \Chi_{(p,\psi)}(\tau,z)  
\\[1em]
&+& \sum_p \frac12 \eE^{\pi\ii \psi} S_{ip} \sum_{\psi=0}^1
\Chi_{(p,\psi)}(\tau,z) \, .
\end{array}\labl{822}
Note that primary fields in the twisted sector have now appeared on the right.
This can be traced back to the fact that we had to split each diagonal field 
in the untwisted sector into two fields with different characters. 
Equation \erf{822} implies the following orbifold $S$-matrix elements:
\be \begin{array}{lll}
S_{(i,\psi),(pq)} &=& S_{ip}S_{iq} \\[1em]
S_{(i,\psi),(j,\chi)} &=& \frac12 S_{ij}^2 \\[1em]
S_{(i,\psi),\Widetilde{(p,\chi)}} &=& \frac12 \eE^{2\pi\ii\psi/2} S_{ip} \, . 
\end{array}\ee

To compute the modular transformation properties of the twisted 
sector, we remark first that 
\be\begin{array}{lll}
\chi_i(\frac{-1+\tau}{2\tau}, \frac z\tau) &=&
T_i \chi_i(\frac{-1-\tau}{2\tau}, \frac z\tau) \\[1em]
&=& \sum_j T_i S_{ij} 
\chi_j(\frac{2\tau}{\tau+1}, \frac{2z}{\tau+1}) \\[1em]
&=& \sum_j T_i S_{ij} T_j^2
\chi_j(\frac{-2}{\tau+1}, \frac{2z}{\tau+1}) \\[1em]
&=& \sum_{jl} T_i S_{ij} T_j^2 S_{jl}
\chi_l(\frac{\tau+1}2, z)  \\[1em]
&=& \sum_{l} (TST^2S)_{il}
\chi_l(\frac{\tau+1}2, z) \, . \end{array}\ee
Using this equation, we find the modular transformation of the characters 
\erf{422} of the primary fields in the twisted sector 
\be\begin{array}{lll}
\Chi_{\Widetilde{(k,\psi)}} (-\frac1\tau) &=& 
\sum_p \frac12 S_{kp}\chi_p(2\tau, 2z) + \eE^{\pi\ii\psi} \frac12 T_k^{-1/2}
(TST^2S)_{kl} \chi_l(\frac{\tau+1}2,z) \\[1em]
&=& \sum_p \frac12 S_{kp} \sum_{\chi=0}^1 \eE^{\ii\pi\chi} 
   \Chi_{(p,\chi)}(\tau,z) 
+ \eE^{\pi\ii\psi} \frac12 \sum_p (T^{1/2} S T^2 S T^{1/2})_{kp} 
\sum_{\chi=0}^1 \eE^{\ii\pi\chi} \Chi_{\Widetilde{(p,\chi)}}(\tau,z) \, . 
\end{array}\ee
It is useful to introduce the symmetric, unitary matrix $P$ as 
\be P \equiv T^{1/2} S T^2 S T^{1/2} \, \, , \, \, P^2 = S^2  \labl{P}
and then we can write our result as 
\be \begin{array}{lll}
S_{\Widetilde{(k,\psi)},(pq)} &=& 0 \\[1em]
S_{\Widetilde{(k,\psi)},(j,\chi)} &=& \eE^{\pi\ii\chi}\frac12 S_{kj} \\[1em]
S_{\Widetilde{(k,\psi)},\Widetilde{(q,\chi)}} 
&=& \eE^{\pi\ii(\psi+\chi)}\frac12 P_{pq}
\end{array}\ee
This completes the computation of the modular matrices of the orbifold for
induction order $\lambdam=2$.

The matrix $P$ introduced in \erf P has also arisen in the 
theory of open and unoriented strings \cite{bisa}, where it describes the 
transition from horizontal to vertical time on a M\"obius strip. This is
more than a coincidence: Given a closed string theory, the 
construction of the open string theory can be thought of as a `parameter
space orbifold' in the sense that its worldsheet can be obtained from an 
oriented closed Riemann surface by dividing out an {\em anti}conformal 
involution. This amounts to modding out the $\zet_2$ symmetry which 
permutes the chiral and the anti-chiral algebra of a left-right symmetric
theory.  It is therefore not surprising that we encounter the same quantities 
in our orbifolds as in the construction of open strings. However, in the case 
of the $\zet_2$ permutation orbifold, the
interpretation is quite different: In our case the orbifold is
associated to a transformation (see Section 3) which is still locally
conformal. Moreover, we still have left and right movers
and the orbifolds we consider are still defined only on closed oriented 
Riemann surfaces.

\subsection{Checks}

In the following checks on the modular matrices of the orbifold, we have
assumed the familiar properties of the modular matrices of the mother theory.
It is obvious that the resulting modular matrix $S$ for the orbifold theory
is symmetric, $S=S^t$. Moreover, a straightforward calculation shows that 
the $S$-matrix of the orbifold theory is unitary, given the unitarity of the
original $S$-matrix. One can also show that the square of the new $S$-matrix 
is a permutation (of order two) of the primary fields, as it should be.
Explicit computation shows that $(ST)^3= S^2$, i.e.\ $S$ together with the 
modular matrix $T$ gives a (projective) \rep\ of the modular group.
We have also checked that the $S$-matrix of the orbifold theory obeys
the required relations $ S_{0,k} \geq S_{00} > 0 $, so that the quantum 
dimensions have the usual properties. 

As another check of our result we have considered the $\zet_2$-orbifold of the 
tensor product of two Ising models: The critical
Ising model corresponds to a free massless fermion, and the twisted sectors
of the Ising orbifolds at $\ch=\lambdam/2$ have
been discussed explicitly in Subsection 3.5. For $\lambdam=2$, 
the cyclic orbifold of the tensor product of two Ising models is a \cft\ with
Virasoro central charge $\ch=1$, and this cyclic orbifold can be identified 
\cite{klsC} as the rational $\zet_2$ orbifold of the free boson with 13 
primary fields.  Our formulas correctly reproduce the modular matrix $S$ 
\cite{dvvv} of the $\ch=1$ orbifold.

\subsection{Orbifold fusion rules}

We can now use the Verlinde formula \erf{verl} to obtain the fusion rules
of the orbifold theory. It is straightforward to check that one 
obtains the expected twist selection rules: The fusion product of two 
fields in the untwisted sector contains only fields in the untwisted sector, 
the fusion of two fields in the twisted sector contains only fields in the 
untwisted sector, and the fusion product of a field in the untwisted sector
with one in the twisted sector of the orbifold lies in the twisted sector. 

The non-vanishing fusion coefficients in the untwisted sector can be computed 
explicitly from the Verlinde formula \erf{verl} and expressed in terms of the 
fusion coefficients $\caln_{ijk}$ of the mother theory (which are non-negative 
integers). The results are:
\be\begin{array}{lll}
\caln_{(ij)(kl)(qp)} &=& \caln_{ikp}\caln_{jlq}+\caln_{ikq}\caln_{jlp}
+\caln_{ilp}\caln_{jkq} + \caln_{ilq}\caln_{jkp} \\[1em]
\caln_{(ij)(pq)(r,\psi)} &=& \caln_{ipr}\caln_{jqr} + 
\caln_{iqr}\caln_{jpr} \\[1em]
\caln_{(ij)(p,\psi)(q,\phi)} &=& \caln_{ipq}\caln_{jpq} \\[1em]
\caln_{(i,\psi)(j,\phi)(k,\chi)} &=& \frac12 \caln_{ijk}
(\caln_{ijk} + \eE^{\pi\ii(\psi+\phi+\chi)} )  \, . 
\end{array}\ee
The first three expressions are manifestly non-negative integers; moreover,
the last expression is also a non-negative integer, since the factor
$\exp(\ii\pi(\psi+\phi+\chi))$ can only take the values $-1$ and $1$ 
when $\lambdam=2$.

For the twisted sector, we obtain the following results for the non-vanishing
fusion coefficients:
\be \caln_{(ij) \Widetilde{(p,\psi)} \Widetilde{(q,\chi)}}
= \sum_s \frac{S_{is}S_{js}S_{ps}S_{qs}}{S_{0s}^2}
= \sum_s \caln_{ijs} \caln_{pqs^+} \, , \labl{433}
where the sums are over primary fields $s$ in the mother theory and 
$s^+$ denotes the primary field that is conjugate to $s$. In the last step
we used the Verlinde formula and the unitarity of the $S$-matrix of the
mother theory. From the last expression in \erf{433} we see that the number 
$\caln_{(ij) \Widetilde{(p,\psi)} \Widetilde{(q,\chi)}}$ is just
the dimension of the space of conformal blocks of the mother theory
for the four-point
function with the primary fields $i,j,p$ and $q$ as the insertions. 
Finally, we obtain
\be 
\caln_{(i\phi) \Widetilde{(j,\psi)} \Widetilde{(k,\chi)}}
= \frac12 \sum_s \frac{S_{is}^2S_{js}S_{ks}}{S_{0s}^2} +
\frac12 \eE^{\pi\ii(\phi+\psi+\chi)} \sum_s 
\frac{S_{is}P_{js} P_{ks}}{S_{0s}} \, . \labl{731}
The sum in the first term is the dimension of a space of conformal 
blocks of four point functions of the mother theory, this time with 
insertions $i,i,j$ and $k$. The second term in \erf{731} has also arisen
in open string theory, and it has been argued \cite{prss} that this
term is an integer as well, since it
describes oriented fusion rules in front of a crosscap. These results 
imply that the fusion coefficients \erf{731} of the orbifold theory are
integer or half-integer. Since 
$\caln_{(i\phi) \Widetilde{(j,\psi)} \Widetilde{(k,\chi)}}$ is a fusion
coefficient of the orbifold theory, and therefore must be a non-negative
integer, we are led to the conjecture that the expression in \erf{731} is 
a non-negative integer in any rational \cft.

\sect{Discussion}

Many of the aspects of cyclic orbifolds presented in this paper 
can be generalized to arbitrary values of $\lambdam$ and even
further to orbifolds of tensor products of a \cft\ by the full permutation 
group $S_n$ or any of its subgroups.
A complete treatment of these issues is, however, beyond the scope of this
paper, and we briefly sketch here only some aspects of these extensions.
\\[2em]
\underline{\bf The new fixed point problem}\\[2em]
The first generalization of our results is to cyclic orbifolds where 
$\lambdam>2$ is still prime. In this case, however, the comparison of \erf{719} 
with the modular invariant partition function of the twisted sector derived 
in \cite{klsC} 
showed that one must include $\lambdam-1$ copies of the characters \erf{422}
of the twisted sector. Sets of $\lambdam-1$ identical characters were also
found in the untwisted sector.  In this situation one has to face the so-called 
fixed point problem \cite{scya6}, where one cannot read off the modular
matrix $S$ from the modular transformation properties of the characters as
functions of $\tau$ and $z$, as we have done for $\lambdam=2$. 
(Similar problems occur in the description of coset \cfts\ \cite{fusS4}
and of integer spin simple current modular invariants \cite{fusS6}.)
\\[2em]
\underline{\bf Copies of the twisted sector on the sphere}\\[2em]
Another implication of this observation is 
that we must include $\lambdam-1$ copies of the twisted
sector of the orbifold on the sphere, and in what follows, we discuss the
interpretation of these copies.

For brevity, we discuss only $\lambdam=3$ and we choose the simplest case,
where the chiral algebra of the mother theory is an affine \lie\ at level $k$.
In this case, the chiral algebra $\cala$ of the tensor product 
theory decomposes into three subspaces which transform under
cyclic permutations with the same eigenvalue. The invariant subspace $\calao$ 
is the subalgebra spanned by the operators that are
invariant under cyclic permutations
\be J_a(z)\otimes\bfe\otimes\bfe + \bfe\otimes J_a(z)\otimes \bfe
+ \bfe\otimes \bfe \otimes J_a(z) \, \ee
which span an affine \lie\ at level $3k$. Again, the analogue of equation 
\erf{720} implies
that the abstract algebra generated by these operators should be 
represented in the twisted sector of the orbifold by integer moded 
operators, 
\be J_a(z)\otimes\bfe\otimes\bfe + \bfe\otimes J_a(z)\otimes\bfe 
+ \bfe\otimes \bfe \otimes J_a(z) \quad \to \quad
\Jahr a 0 z \, . \ee
The other two subspaces $\calap$ and $\calam$ are spanned by the 
linear combinations
\be \begin{array}{ll}\calap\, : \quad & J_a(z)\otimes\bfe\otimes\bfe 
+ \eE^{2\pi\ii/3} \bfe\otimes J_a(z)\otimes \bfe
+ \eE^{-2\pi\ii/3} \bfe\otimes \bfe \otimes J_a(z) \, \\[1em]
\calam\, : \quad & J_a(z)\otimes\bfe\otimes\bfe 
+ \eE^{-2\pi\ii/3} \bfe\otimes J_a(z)\otimes \bfe
+ \eE^{2\pi\ii/3} \bfe\otimes \bfe \otimes J_a(z) \,.\end{array} \ee
These two subspaces both transform in the adjoint \rep\ with respect to
$\calao$.

It is clear that the representatives of $\calap$ and $\calam$ in the
twisted sector appear as the orbifold currents $\Jahr a 1 z$ and
$\Jahr a 2 z$. Since there is no principle by which to decide the precise
identification of $\cala^{(\pm)}$ with the twisted currents, both possibilities
must occur,
\be \begin{array}{lll}
\mbox{sector 1:}\quad & \calap \to \Jahr a 1 z \quad &
\calam \to \Jahr a 2 z \\[1em]
\mbox{sector 2:}\quad & \calam \to \Jahr a 1 z \quad &
\calap \to \Jahr a 2 z \end{array}\ee
and this is the interpretation of the two identical twisted sectors at 
$\lambdam=3$. This observation is easily generalized to interpret the
multiplicity $\lambdam-1$ for arbitrary prime $\lambdam$.

To study cyclic permutations for arbitrary $\lambdam$, one
should combine the fixed point resolution for $\lambdam$ prime with an
analysis that parallels the one given for the untwisted sector
in Subsection 4.2 (which takes into account the different 
stabilizers in the theory). 
\\[2em]
\underline{\bf Non-abelian permutation orbifolds}\\[2em]
We also expect that our analysis will be useful in the construction of
arbitrary permutation orbifolds, since it is known \cite{dmvv} that the
twisted sectors of these more general orbifolds are composed entirely of
symmetrized combinations of the sectors of cyclic orbifolds.

In this connection, we also remark that the orbifold induction procedure
includes cyclic orbifolds in which the mother theory is itself a cyclic
orbifold or even a general orbifold. It would be interesting to clarify the 
relation of 
non-abelian permutation orbifolds to these more complicated orbifolds.
\\[2em]
\underline{\bf Generalized coset constructions}\\[2em]
Using a construction related to our orbifold induction procedure, Kac and
Wakimoto \cite{kawa5} and Bouwknegt \cite{bouw6} have proposed a new
coset construction, which reads 
\be \begin{array}{lll}
\hat T_{(\g;\h)} &=& \drac{L^{ab}_{\g}(k)}{\lambdam} \sum_{r=0}^{\lambdam-1}
:\Jahhr a r \Jahhr b {-r}: \,\,\, - \, \, 
L^{ab}_{\h}(\lambdam k_\h) : \Jahhr a 0 \Jahhr b 0: 
\,\,+\,\, \drac{\ch_\g(k)}{24\lambdam z^2}(1-\drac1{\lambdam^2}) \\[2em]
\ch_{(\g;\h)} &=& \ch_\g(k) - c_\h(\lambdam k_\h) \,  
\end{array}\labl{constr}
when written in terms of the orbifold currents $\Jahhr a r$
at level $\hat k=\lambdam k$. Here $\ch_\g(k)$
is the orbifold affine-Sugawara central charge $\ch_\g$ in \erf{3.51new}
and
$c_{\h}(k_\h)$ is the affine-Sugawara central charge for level $k_\h$ of
$\h\subset\g$. The construction
\erf{constr} is a straightforward generalization of the one in Refs.\ 
\cite{kawa5,bouw6}, which considered only the case $\h=\g$. 

The results of our paper point to a conjecture that the constructions 
\erf{constr} are in fact an orbifold induction procedure for the twisted
sectors of another type of orbifold:
Consider the coset \cft\ defined by embedding $\h$ diagonally into 
the tensor product of $\lambdam$ copies of the affine \lie\ $\g$. It
is possible to show that the coset theory has a residual action of 
the cyclic permutation symmetry $\zet_\lambdam$ of the ambient algebra. 
Then it is natural to conjecture that the new coset constructions \erf{constr}
give the twisted sectors of the orbifold theories obtained by modding out 
this $\zet_\lambdam$ symmetry. 

There is some support for this conjecture, at least in the case when $\h=\g$: 
Generally, the chiral algebra of the coset theory $(\g\oplus\g)/\g$ is 
a Casimir $\rm W$ algebra \cite{bbss,bbss2},
which includes a Virasoro subalgebra.
On the other hand, it was observed in Ref.\ \cite{bouw6} that for
$\ch<1$ the primary fields of the new coset construction are typically
those that are not present in the theory with $\rm W$ symmetry.
This is exactly the role of twisted sectors, according to our general
discussion of orbifolds in Subsection 3.8.
Indeed, the extinction by orbifoldization of a $\Wt$
symmetry was noted explicitly for the orbifold $\Wt$ algebra 
discussed in Subsections 2.5 and 2.7.
\\[2em]
\underline{\bf Generalized Virasoro master equation}\\[2em]
It is clear that the affine-Virasoro construction \cite{haki,mprst,Hkoc}
can now be generalized to
include the orbifold currents $\Jahhr a r, r=0\ldots\lambdam-1$ which
satisfy the orbifold affine algebra \erf{2.1}. The general 
quadratic ansatz for the chiral stress tensor has the form
\be \hat T({\cal L})  
= \sum_{r=0}^{\lambdam-1} \call^{ab}_r :\Jahhr a r \Jahhr b {-r}:
\, + \, \mbox{$c$-number term} \, \ee
where the coefficients $\call^{ab}_r, r=0\ldots\lambdam-1$ are to be determined
by a generalized Virasoro master equation. The solutions of this system will
therefore include \\
$\bullet$~~ The affine-Virasoro construction itself, when $\call^{ab}_r
= L^{ab} \delta_{r,0}$ and $L^{ab}$ is a solution of the Virasoro
master equation at level $\hat k$, \\
$\bullet$~~ The orbifold integral Virasoro subalgebras $\hat T^{(0)}$ of 
this paper (see eq. \ref{3.67new})
            when $\call^{ab}_r=\frac1\lambdam L^{ab}$,  \\
$\bullet$~~ The new coset constructions \erf{constr}, \\
as well as a presumably large class of Virasoro operators of (sectors of) 
new \cfts, beyond these listed above.
\vskip4em

  \newcommand{\wb}{\,\linebreak[0]} \def\wB {$\,$\wb}
  \newcommand{\Bi}[1]    {\bibitem{#1}}
  \newcommand{\Erra}[3]  {\,[{\em ibid.}\ {#1} ({#2}) {#3}, {\em Erratum}]}
  \newcommand{\BOOK}[4]  {{\em #1\/} ({#2}, {#3} {#4})}
  \newcommand{\vypf}[5]  {{#1} [FS{#2}] ({#3}) {#4}}
  \renewcommand{\J}[5]   {{#1} {#2} ({#3}) {#4} }
  \newcommand{\Prep}[2]  {{\sl #2}, preprint {#1}}
 \def\acam  {Acta\wB Appl.\wb Math.}
 \def\anop  {Ann.\wb Phys.}
 \def\bams  {Bull.\wb Amer.\wb Math.\wb Soc.}
 \def\coma  {Con\-temp.\wb Math.}
 \def\foph  {Fortschr.\wb Phys.}
 \def\fuaa  {Funct.\wb Anal.\wb Appl.}
 \def\hepa  {Helv.\wb Phys.\wB Acta}
 \def\ijmp  {Int.\wb J.\wb Mod.\wb Phys.\ A}
 \def\inma  {Invent.\wb math.}
 \def\imrn  {Int.\wb Math.\wb Res.\wb Notices}
 \def\jodg  {J.\wb Diff.\wb Geom.}
 \def\jopa  {J.\wb Phys.\ A}
 \def\jomp  {J.\wb Math.\wb Phys.}
 \def\npbF  {Nucl.\wb Phys.\ B\vypf}
 \def\npbp  {Nucl.\wb Phys.\ B (Proc.\wb Suppl.)}
 \def\nuci  {Nuovo\wB Cim.}
 \def\nupb  {Nucl.\wb Phys.\ B}
 \def\phlb  {Phys.\wb Lett.\ B}
 \def\pnas  {Proc.\wb Natl.\wb Acad.\wb Sci.\wb USA}
 \def\prep  {Phys.\wb Rep.}
 \def\comp  {Com\-mun.\wb Math.\wb Phys.}
 \def\lemp  {Lett.\wb Math.\wb Phys.}
 \def\phrd  {Phys.\wb Rev.\ D}
 \def\mpla  {Mod.\wb Phys.\wb Lett.\ A}
 \def\duke  {Duke\wB Math.\wb J.}

 \def\A       {Algebra}
 \def\alg     {algebra}
 \def\Be     {{Berlin}}
 \def\BIR    {{Birk\-h\"au\-ser}}
 \def\Ca     {{Cambridge}}
 \def\CUP    {{Cambridge University Press}}
 \def\furu    {fusion rule}
 \def\GB     {{Gordon and Breach}}
 \newcommand{\inBO}[7]  {in:\ {\em #1}, {#2}\ ({#3}, {#4} {#5}),  p.\ {#6}}
 \def\Infdim  {Infinite-dimensional}
 \def\NY     {{New York}}
 \def\nn      {$N=2$ }
 \def\Q       {Quantum\ }
 \def\qg      {quantum group}
 \def\Rep     {Representation}
 \def\SV     {{Sprin\-ger Verlag}}
 \def\syms    {sym\-me\-tries}
 \def\wzw     {WZW\ }

\bigskip\bigskip
\small
\noindent{\bf Acknowledgement.} \ We are grateful to P.\ Bouwknegt, 
J.\ de Boer, C.J.\ Efthimiou, J.\ Fuchs, P.\ Ho\v{r}ava, J.\ Lepowsky, 
H.\ Ooguri and A.\ Sagnotti for helpful discussions. \\
M.B.H.\ and C.S.\ were supported in part by the Director, Office of Energy 
Research,
Office of Basic Energy Sciences, of the U.S.\ Department of Energy under
Contract DE-AC03-76F00098 and in part by the National Science Foundation
under grant PHY95-14797.

\bigskip\bigskip\noindent
After submission of this paper, we were informed that the orbifold Virasoro
algebra has also appeared \cite{zamo6,crss2,fesu,apef}
in the context of \cft\ on $\zet_\lambdam$-symmetric higher-genus Riemann 
surfaces. Moreover, the integral affine subalgebras and vertex
operators closely related to our higher-level vertex operator construction
have been used in the construction of higher-level standard modules of
affine SU(2) \cite{lePr}.


\end{document}